\documentclass[twocolumn,aps,superscriptaddress,showpacs,nofootinbib,floatfix]{revtex4}
\usepackage{epsfig,bm,feynmf}
\usepackage{graphics}
\usepackage{amsmath}
\usepackage[normalem]{ulem}  
\usepackage[dvips]{color} 

\renewcommand\sout{\bgroup \color{red} \ULdepth=-.5ex \ULset}
\begin{document}


\title{Open charm and dileptons from relativistic heavy-ion collisions}


\author{Taesoo Song}\email{taesoo.song@theo.physik.uni-giessen.de}
\affiliation{Institut f\"{u}r Theoretische Physik, Universit\"{a}t
Gie\ss en, Germany}

\author{Wolfgang Cassing}
\affiliation{Institut f\"{u}r Theoretische Physik, Universit\"{a}t
Gie\ss en, Germany}

\author{Pierre Moreau}
\affiliation{Institute for Theoretical Physics, Johann Wolfgang
Goethe Universit\"{a}t, Frankfurt am Main, Germany}

\author{Elena Bratkovskaya}
\affiliation{Institute for Theoretical Physics, Johann Wolfgang
Goethe Universit\"{a}t, Frankfurt am Main, Germany}
\affiliation{GSI
Helmholtzzentrum f\"{u}r Schwerionenforschung GmbH, Planckstrasse 1,
64291 Darmstadt, Germany}


\begin{abstract}
Dileptons are considered as one of the cleanest signals of the quark-gluon plasma (QGP),
however, the QGP radiation is masked by many 'background'  sources from either hadronic decays
or semileptonic decays from correlated charm pairs.
In this study we investigate the relative contribution of these channels in heavy-ion collisions from
$\sqrt{s_{\rm NN}}=$ 8 GeV to 5 TeV with a focus on the competition between
the thermal QGP radiation and the semileptonic decays from correlated $D-$meson pairs.
As a 'tool' we employ the parton-hadron-string dynamics (PHSD) transport
approach to study dilepton spectra in Pb+Pb (Au+Au) collisions in a wide energy range
incorporating for the first time a fully microscopic treatment of the charm dynamics and their semileptonic decays. We find that the  dileptons from correlated $D-$meson decays
dominate the 'thermal' radiation from the QGP in central Pb+Pb collisions
at the intermediate masses (1.2 GeV $< M <$ 3 GeV) for
$\sqrt{s_{\rm NN}} > $ 40 GeV, while for $\sqrt{s_{\rm NN}}=$ 8 to 20 GeV
the contribution from $D,{\bar D}$ decays to the
intermediate mass dilepton spectra is subleading such that one should
observe a rather clear signal from the QGP radiation.
We, furthermore, study the $p_T$-spectra and the $R_{AA}(p_T)$
of single electrons at different energies as well
as the excitation function of the inverse slope of the $m_T$- spectra for intermediate-mass dileptons from the
QGP and from charm decays. We find moderate but characteristic changes in the inverse slope parameter for
$\sqrt{s_{\rm NN}} > $ 20 GeV which can be observed experimentally in high statistics data.
Additionally,  we provide detailed predictions for dilepton spectra from Pb+Pb collisions
at $\sqrt{s_{\rm NN}} = $ 5.02 TeV.
\end{abstract}


\maketitle

\section{Introduction}

Relativistic heavy-ion collisions are well suited  to generate hot
and dense matter in the laboratory, although the matter is produced within small space-time
regimes. Whereas in low energy collisions one produces dense nuclear
matter with moderate temperature $T$ and large baryon chemical potential
$\mu_B$, ultra-relativistic collisions at Relativistic Heavy Ion
Collider (RHIC) or Large Hadron Collider (LHC) energies produce
extremely hot matter at small baryon chemical potential. In order to
explore the phase diagram of strongly interacting matter as a
function of $T$ and $\mu_B$ both type of collisions are mandatory.
According to lattice calculations of quantum chromodynamics
(lQCD)~\cite{Bernard:2004je,Aoki:2006we,Bazavov:2011nk}, the phase
transition from hadronic to partonic degrees of freedom (at
vanishing baryon chemical potential $\mu_B$=0) is a crossover. This
phase transition is expected to turn into a first order transition
at a critical point $(T_r, \mu_r)$ in the phase diagram with
increasing baryon chemical potential $\mu_B$. Since this critical
point cannot be  determined theoretically in a reliable way the beam
energy scan (BES) program  at  RHIC  aims to find the
critical point and the phase boundary by gradually decreasing the
collision energy~\cite{Mohanty:2011nm,Kumar:2011us}. Furthermore,
new facilities such as FAIR (Facility for Antiproton and Ion
Research) and NICA (Nuclotron-based Ion Collider  fAcility) are
under construction to explore in particular the intermediate energy
range where one might study also the competition between chiral
symmetry restoration and deconfinement as suggested in Refs.
\cite{Cas16,Palmese}.

Since the partonic phase  in relativistic heavy-ion collisions
appears only for a couple of fm/c, it is quite a  challenge for
experiment to investigate its properties. The heavy flavor mesons
are considered to be promising probes in this search since the
production of heavy flavor requires a large energy-momentum
transfer. Thus it takes place early in the heavy-ion collisions, and
- due to the large energy-momentum transfer - should be described by
perturbative quantum chromodynamics (pQCD). The produced heavy
flavor then interacts with the hot dense matter
(of partonic or hadronic nature) by exchanging energy and momentum.
As a result, the ratio of the measured number of heavy flavors in
heavy-ion collisions to the expected number in the absence of
nuclear or partonic matter is suppressed at high transverse
momentum, and the elliptic flow of heavy flavor is generated by the
interactions in noncentral heavy-ion collisions.  The experimental
data at RHIC and LHC show that the suppression of heavy-flavor
hadrons at high transverse momentum and its elliptic flow $v_2$ are
comparable to those of light
hadrons~\cite{ALICE:2012ab,Abelev:2013lca}. This is a puzzle for
heavy-flavor production and dynamics in relativistic heavy-ion
collisions as pointed out by many groups
~\cite{Moore:2004tg,Zhang:2005ni,Molnar:2006ci,Linnyk:2008hp,Gossiaux:2010yx,Nahrgang:2013saa,He:2011qa,He:2012df,He:2014epa,Uphoff:2011ad,Uphoff:2012gb,Cao:2011et,Greco,Nahrgang:2016lst}
and a subject of intense studies both theoretically and
experimentally. For recent reviews we refer the reader to Refs.
\cite{Andro,Rapp16}.

Furthermore, the electromagnetic emissivity of strongly interacting
matter is a subject of longstanding interest
\cite{Feinb,Shur,Bauer,ChSym} and is explored also in relativistic
nucleus-nucleus collisions, where the photons (and dileptons)
measured experimentally provide a time-integrated picture of the
collision dynamics. Especially dileptons are of particular interest
since their invariant mass provides an additional scale compared to
photons and allows to partly separate the production channels from
the early (possibly partonic) phase with those from the late
hadronic phase. After decades of experimental and theoretical
studies it has become clear that dileptons with invariant masses
below about 1.2 GeV preferentially stem from hadronic decays
providing some glance at the modification of hadron properties in
the dense and hot hadronic medium  (cf. \cite{ChSym,PHSDreview} and
references therein) while the intermediate mass regime 1.2 GeV $< M
<$ 3 GeV should provide information about 'thermal' dileptons from the QGP
($q+{\bar q} \rightarrow e^+ e^-, \ q+\bar{q}\rightarrow g+\gamma^*,
g+q({\bar q}) \rightarrow, \ q({\bar q}) + e^+ e^-$)
as well as the amount of correlated open
charm (semileptonic) decays from early production of $c{\bar c}$
pairs. Whereas at RHIC and LHC energies the background from $D {\bar
D}$ pairs overshines the contribution from the QGP in the
intermediate mass regime \cite{PHSDreview}, one might expect to find
some window in bombarding energy where the partonic sources dominate
since the charm production drops rapidly with decreasing bombarding
energy. In this work we intend to quantify this expectation and to
identify optimal systems for future measurements at FAIR/NICA
or at the RHIC Beam-Energy-Scan (BES) as well as at
the Super Proton Synchrotron (SPS) by the NA61 collaboration.

We recall that previously we have studied the contribution of semileptonic decays
from $D$-mesons to the dilepton spectra at RHIC ($\sqrt{s_{\rm NN}}$ = 200 GeV)
and LHC ($\sqrt{s_{\rm NN}}$ = 2.76 TeV) energies based on
an extended statistical hadronization model \cite{Jaako,Jaako2}.
The charm production in $AA$ collisions was accounted for by scaling the contribution
from $p+p$ collisions with the number of binary $NN$ collisions.
However, in these studies only the semileptonic decays of correlated
(and unscattered) $D {\bar D}$ pairs were considered whereas the contribution
from rescattered $D$ and ${\bar D}$ mesons had been neglected.
Also only hadronic rescattering has been incorporated for the decorrelation
of the produced $D {\bar D}$ pair.
Since these assumptions are too crude to correctly reflect the actual experimental
measurements with their detailed acceptance cuts a fully microscopic reanalysis of the
charm dynamics and charm pair angular correlation is mandatory.

We here employ the microscopic parton-hadron-string dynamics (PHSD) approach, which
differs from the conventional Boltzmann-type models in the
aspect~\cite{Cassing:2009vt} that the degrees-of-freedom for the QGP
phase are off-shell massive strongly-interacting quasi-particles
that generate their own mean-field potential. The masses of the
dynamical quarks and gluons in the QGP are distributed according to
spectral functions whose pole positions and widths, respectively,
are defined by the real and imaginary parts of their self-energies
\cite{PHSDreview}. The partonic propagators and self-energies,
furthermore, are defined in the dynamical quasiparticle model (DQPM)
in which the strong coupling and the self-energies are fitted to
lattice QCD results \cite{Cassing:2008nn}.

We recall that the PHSD approach has successfully described numerous
experimental data in relativistic heavy-ion collisions from the
Alternating Gradient Synchrotron (AGS), SPS, RHIC  to LHC
energies~\cite{Cassing:2009vt,PHSDrhic,Volo,PHSDreview,Eduard}. More
recently, the charm production and propagation has been explicitly
implemented in the PHSD and detailed studies on the charm dynamics
and hadronization/fragmention have been performed at top RHIC and
LHC energies in comparison to the available
data~\cite{Song:2015sfa,Song:2015ykw,Song2016}. In the PHSD approach
the initial charm and anticharm quarks are produced by using the
PYTHIA event generator~\cite{Sjostrand:2006za} which is tuned to the
transverse momentum and rapidity distributions of charm and
anticharm quarks from the Fixed-Order  Next-to-Leading Logarithm
(FONLL) calculations~\cite{Cacciari:2012ny}. The produced charm and
anticharm quarks interact in the QGP with off-shell partons and are
hadronized into $D-$mesons close to the critical energy density
($\sim$ 0.5 GeV/fm$^3$) for the crossover transition either through
fragmentation or coalescence. We stress that the coalescence is a
genuine feature of heavy-ion collisions and does not show up in p+p
interactions. The hadronized $D-$mesons then interact with light
hadrons in the hadronic phase until freeze out and final
semileptonic decay. We have found that the PHSD approach, which has
been applied for charm production in Au+Au collisions at
$\sqrt{s_{\rm NN}}=$200 GeV~\cite{Song:2015sfa} and in Pb+Pb
collisions at $\sqrt{s_{\rm NN}}=$2.76 TeV~\cite{Song:2015ykw},
describes the $R_{\rm AA}$  of $D-$mesons  in reasonable agreement
with the experimental data from the STAR
collaboration~\cite{Adamczyk:2014uip,Tlusty:2012ix} and from the
ALICE collaboration~\cite{Adam:2015sza,Abelev:2014ipa} when
including the initial shadowing effect in the latter case. In this
work we, furthermore, apply the PHSD approach to charm and dilepton
production in relativistic heavy-ion collisions from
$\sqrt{s_{\rm NN}}=$ {8 GeV to 2.76 TeV}, analyse the angular correlation
between the charm quarks or $D$-mesons, respectively, and evaluate
the contribution to the dilepton spectra from their semileptonic
decays.  Furthermore,
we will give predictions for dilepton mass spectra from Pb+Pb collisions
at the top LHC energy of $\sqrt{s_{\rm NN}}=$ 5.02 TeV for low and intermediate invariant masses.

This paper is organized as follows: The production of heavy mesons
in p+p collisions is described in Sec. II and $c{\bar c}$ pair
multiplicities in central Pb+Pb collisions are evaluated within PHSD
as a function of invariant energy. We then present the heavy quark
interactions in the QGP, their hadronization and
hadronic interactions, respectively, in Sec. III as well as the
semileptonic decays of the charm hadrons. Sec. IV is devoted to the
description of the dilepton sources incorporated in the actual PHSD calculations
while in Sec. V we calculate the $R_{AA}$ of single electrons from open charm mesons at midrapidity as a function of
transverse momentum and the modification of the $c{\bar c}$
correlation angle due to the partonic and hadronic interactions in central Pb+Pb collisions from
$\sqrt{s_{\rm NN}}=$ 8 to 200 GeV. We continue with excitation functions for dilepton spectra in these collisions
and investigate separately the contributions from hadronic and partonic sources as well as semi-leptonic decays from open charm. In
Sec. VI we will compare the PHSD calculations for dilepton spectra with experimental data
from $\sqrt{s_{\rm NN}}$ = 19.6 GeV to 2.76 TeV and present predictions for dilepton mass spectra from
Pb+Pb collisions at the top LHC energy of $\sqrt{s_{\rm NN}}=$ 5.02 TeV. Sec. VII closes our study
with a summary while Appendices A and B include the details of the partonic production channels for lepton pairs as well as an examination of the uncertainties in the charm cross section and the effects of experimental cuts on the dilepton spectra.

\section{Charm pairs from p+p collisions}\label{pp}

\begin{figure} [h]
\centerline{
\includegraphics[width=8.6 cm]{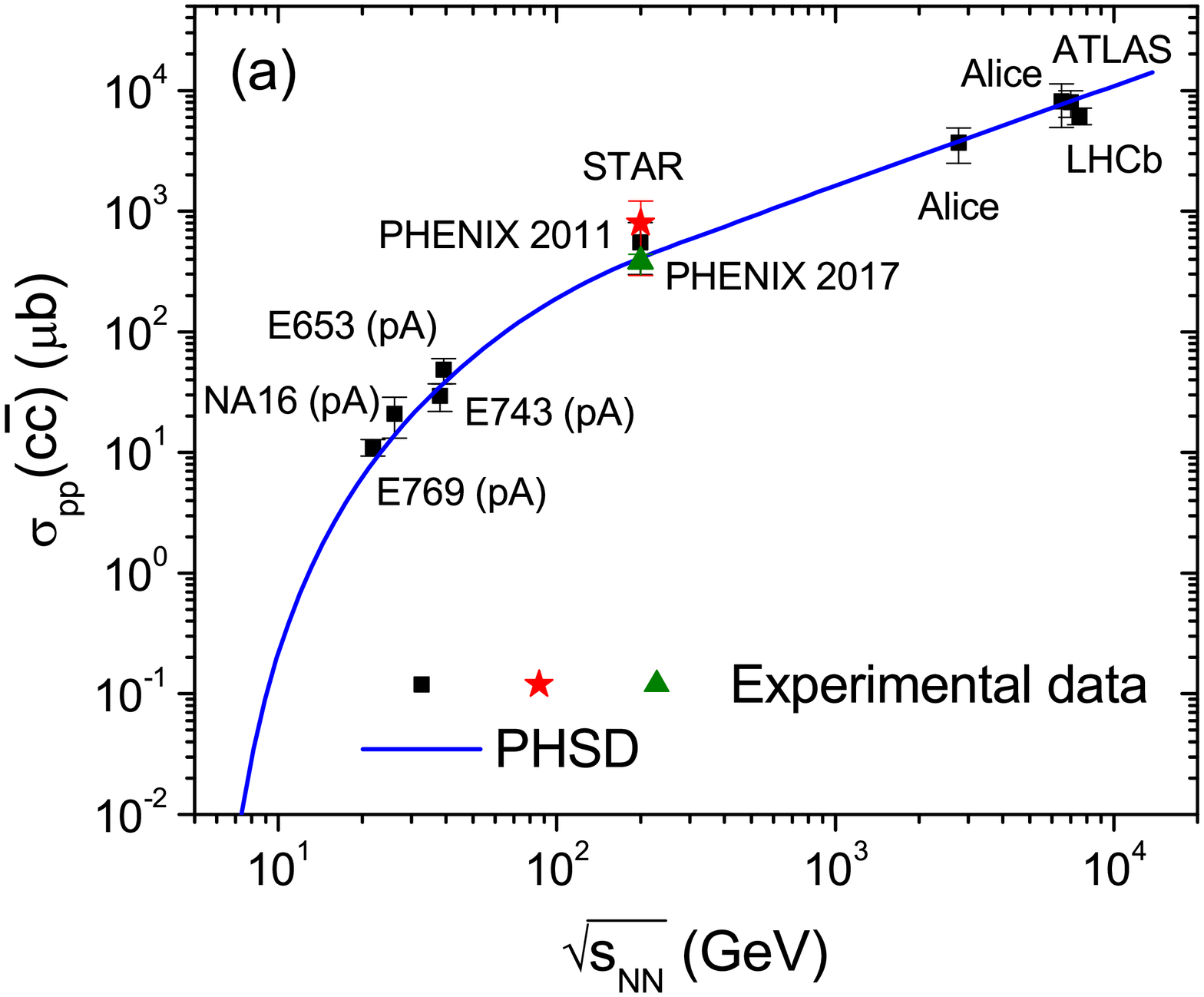}}
\centerline{
\includegraphics[width=8.6 cm]{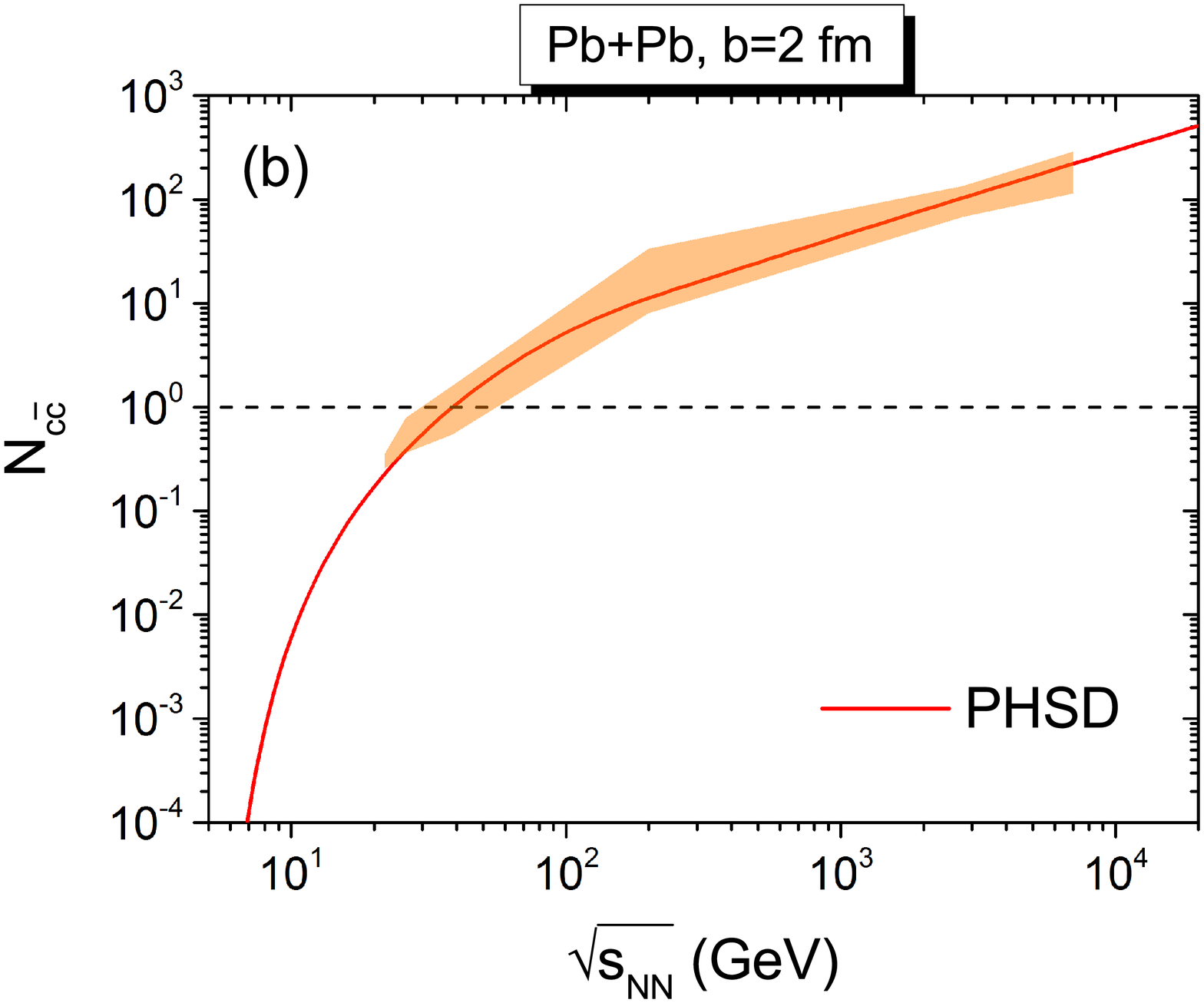}}
\caption{a) The $c{\bar c}$ pair cross section in p+p
reactions as a function of the invariant energy $\sqrt{s_{\rm NN}}$ as implemented in PHSD. The
symbols denote experimental data from Refs. \cite{delValle:2011ex,Adamczyk:2012af,Adare:2017caq}. b) The number of primary
$c{\bar c}$ pairs in Pb+Pb collisions at b=2 fm as a function of
$\sqrt{s_{\rm NN}}$. The shaded area in (b) shows the uncertainty
in the number of $c{\bar c}$ pairs due to the uncertainty in the charm production cross section in p+p collisions.} \label{fig1}
\end{figure}

As pointed out in the Introduction the charm  quark ($c {\bar c}$) pairs are
produced through initial hard nucleon-nucleon scattering in
relativistic heavy-ion collisions. We employ the PYTHIA event
generator to produce the heavy-quark pairs and modify their
transverse momentum and rapidity such that they are similar to those
from the FONLL calculations at RHIC and LHC energies (cf. Ref.
\cite{Song2016}). At SPS and lower energies we do not employ any
modification of the PYTHIA results. Fig.
\ref{fig1} a) {shows the charm production cross section for p+p collisions (as implemented in PHSD)}
as a function of the invariant energy
$\sqrt{s_{\rm NN}}$ {which is fitted to a wide range of experimental data.}
{We can see} a rather fast drop of the $c{\bar c}$ cross
section with decreasing energy especially close to the threshold energy for charm-pair production. Note, however, that
the data show an uncertainty of about a factor of two which implies a corresponding uncertainty
in the following PHSD calculations.

\subsection{Multiplicities for $c{\bar c}$ pairs in central Pb+Pb reactions}
We recall that in heavy-ion reactions the number of $c{\bar c}$
pairs produced is approximately given by the number of binary
nucleon-nucleon collisions $N_{bin}(b)$ (at given impact parameter
$b$) times the probability to produce a $c{\bar c}$ pair in an
inelastic nucleon-nucleon collision at given $\sqrt{s_{\rm NN}}$ which
is the ratio of the $c{\bar c}$ cross section to the
inelastic $N+N$ cross section. The scaling of the $c{\bar c}$
multiplicity with the number of binary $N+N$ collisions is rather
well reproduced in actual PHSD calculations where additionally the
smearing of $\sqrt{s_{\rm NN}}$ by Fermi motion is taken into account as
well as fluctuations in the number of binary nucleon-nucleon
collisions $N_{bin}(b)$ on an event by event basis. The corresponding
PHSD results for Pb+Pb collisions at b= 2 fm are displayed in Fig. \ref{fig1} b) as a function of
$\sqrt{s_{\rm NN}}$ and
demonstrate that the average $c{\bar c}$ pair multiplicity in
central collisions is far below unity at SPS and FAIR/NICA energies. In this case
we may gate in the PHSD calculations on events with a single $
c{\bar c}$ pair - selected by Monte-Carlo from the number of possible
binary $N+N$ reactions - and follow the dynamics of the charm quarks
throughout the time evolution in PHSD, i.e. partonic scattering,
hadronization by coalescence or fragmentation, and final hadronic
rescattering of charmed mesons and baryons (see below). At the end
all observables have to be multiplied by the probability for the
charm event as illustrated in Fig. \ref{fig1} b). The shaded area in Fig.
\ref{fig1} b) shows the uncertainty in the number of $c{\bar c}$ pairs due to the uncertainty
of the charm cross section in p+p collisions (cf. Fig. \ref{fig1} a)).
Note that for $\sqrt{s_{\rm NN}} <$ 20 GeV no data are available and the number of $c{\bar c}$ pairs
entirely stem from a parameterized function which takes into account the phase space of final states.

\subsection{Fragmentation of charm and bottom in p+p collisions}
The produced charm and bottom quarks in hard nucleon-nucleon
collisions are hadronized by emitting soft gluons, which is denoted
by `fragmentation'. As in Ref. \cite{Song:2015sfa} we use the
fragmentation function of Peterson which reads
as~\cite{Peterson:1982ak}
\begin{eqnarray}
D_Q^H(z)\sim \frac{1}{z[1-1/z-\epsilon_Q/(1-z)]^2},
\end{eqnarray}
where $z$ is the momentum fraction of the hadron $H$ fragmented from
the heavy quark $Q$ while $\epsilon_Q$ is a fitting parameter which
is taken to be $\epsilon_Q$ = 0.01 for charm~\cite{Song:2015sfa} {and 0.004 for bottom~\cite{Song:2015ykw}. We note that the fragmentation function is applied only for the transverse momentum of the hadron while the rapidity is assumed to be the same as before the fragmentation.}
The chemical fractions of the charm quark decay into
$D^+,~D^0,~D^{*+},~D^{*0},~D_s$, and $\Lambda_c$ are  taken to be
14.9, 15.3, 23.8, 24.3, 10.1, and 8.7
\%~\cite{Gladilin:1999pj,Chekanov:2007ch,Abelev:2012vra,Song:2015ykw},
respectively{, and those of the bottom quark decay  into
$B^-,~\bar{B^0},~\bar{B^0}_s$, and $\Lambda_b$ are 39.9, 39.9, 11,
and 9.2 \%~\cite{DELPHI:2011aa}}.  After the momentum and the species of the fragmented
particle are decided by  Monte Carlo, the energy of the fragmented
particle is adjusted to be on-shell. Furthermore, the $D^*$ mesons
first decay into $D+\pi$ or $D+\gamma$, and then the $D-$ mesons
produce single electrons through the semileptonic
decay~\cite{Agashe:2014kda}, which is evaluated within PYTHIA.

\section{Heavy quark dynamics in A+A collisions}
We here briefly recall the various interactions of charm quarks (or
charm hadrons) in the partonic (hadronic) medium as introduced in
Ref. \cite{Song2016}.

\subsection{Heavy-quark interactions in the QGP}\label{QGP}
In PHSD the baryon-baryon and baryon-meson collisions at high-energy
produce strings. If the local energy density is above the critical
energy density ($\sim$ 0.5 GeV/fm$^3$), the strings melt into quarks
and antiquarks with masses  determined by the temperature-dependent
spectral functions from the DQPM~\cite{Cassing:2008nn}. Massive
gluons are formed through flavor-neutral quark and antiquark fusion
in line with the DQPM. In contrast to normal elastic scattering,
off-shell partons may change their mass after the elastic scattering
according to the local temperature $T$ in the cell (or local
space-time volume) where the scattering happens. This automatically
updates the parton masses as the  hot and dense  matter expands,
i.e. the local temperature decreases with time. The same holds true
for the reaction chain from gluon decay to quark+antiquark ($g
\rightarrow q + {\bar q}$) and the inverse reaction ($q + {\bar q}
\rightarrow g$) following detailed balance. The local temperature is
determined from the local energy density in the rest frame of the
cell by employing the lattice QCD equation of state from Ref.
\cite{Aoki:2009sc}.

Due to the finite spectral width of the partonic degrees-of-freedom,
the parton spectral function has time-like as well as space-like
parts. The time-like partons propagate in space-time within the
light-cone while the space-like components are attributed to a
scalar potential energy density~\cite{Cassing:2009vt}. The gradient
of the potential energy density with respect to the scalar density
generates a repulsive force in relativistic heavy-ion collisions and
plays an essential role in reproducing experimental flow data and
transverse momentum spectra of hadrons with light quarks (see Ref.
\cite{PHSDreview} for a review). For charm quarks we assume in this
study that the heavy quark has a constant (on-shell) mass: the charm
quark mass is taken to be 1.5 GeV, however, the light
quarks/antiquarks as well as gluons are treated fully off-shell.

The heavy quarks and antiquarks produced in early hard collisions -
as described above - interact with the dressed lighter off-shell
partons in the QGP. The cross sections for the heavy-quark
scattering with massive off-shell partons have been calculated by
considering explicitly the mass spectra of the final-state particles
in Refs.~\cite{Berrehrah:2013mua,Berrehrah:2014kba}. The elastic
scattering of heavy quarks in the QGP is treated by including the
non-perturbative effects of the strongly interacting quark-gluon
plasma (sQGP) constituents, i.e. the temperature-dependent coupling
$g(T/T_c)$ which rises close to $T_c$, the multiple scattering etc.
The multiple strong interactions of quarks and gluons in the sQGP
are encoded in their effective propagators with broad spectral
functions (imaginary parts). As pointed out above, the effective
propagators, which can be interpreted as resummed propagators in a
hot and dense QCD environment, have been extracted from lattice data
in the scope of the DQPM~\cite{Cassing:2008nn}. We recall that the
divergence encountered in the $t$-channel scattering is cured
self-consistently, since the infrared regulator is given by the
finite DQPM gluon mass and width. For further details we refer the
reader to Refs.~\cite{Berrehrah:2013mua,Berrehrah:2014kba}.

We recall that charm interactions in the QGP -- as described by the
DQPM charm scattering cross sections -- differ substantially form
the pQCD scenario, however, the spacial diffusion constant for charm
quarks $D_s(T)$ is consistent with the lQCD data
\cite{Song:2015ykw,Berrehrah:2016vzw} within errorbars.

\subsection{Heavy-quark hadronization}\label{hadronization}
The heavy-quark hadronization in nucleus-nucleus collisions is
realized via 'dynamical coalescence' and fragmentation. Here
`dynamical coalescence' means that the probability to find a
coalescence partner is {calculated from the Wigner density in coordinate and momentum space and the coalescence is realized } by Monte Carlo in the vicinity of the
critical energy density $0.4\le \epsilon \le 0.75$ GeV/fm$^3$ as
described in Ref. \cite{Song2016}. We note that such a dynamical
realization of heavy-quark coalescence is in line with the dynamical
hadronization of light quarks in the PHSD. Summing up the
coalescence probabilities from all candidates, whether the heavy
quark or heavy antiquark hadronizes by coalescence or not, and which
quark or antiquark among the candidates will be the coalescence
partner, is decided by Monte Carlo. If a random number is above the
sum of the coalescence probabilities, it is tried again in the next
time step till the local energy density is lower than 0.4 $\rm
GeV/fm^3$. The heavy quark or heavy antiquark, which does not
succeed to hadronize by coalescence throughout the expansion phase
of the partonic subsystem, then hadronizes through fragmentation as
in p+p collisions. We recall that charm quarks with low transverse
momenta $p_T$ dominantly hadronize {by coalescence} while those with large $p_T$
undergo fragmentation \cite{Song2016}.

\subsection{Interactions of charm mesons with the hadronic
medium}\label{hg} After the hadronization of heavy quarks and their
subsequent decay into $D, D^*$ mesons, the final
stage of the evolution concerns the interaction of these states with
the hadrons forming the expanding bulk medium. A realistic
description of the hadron-hadron scattering
---potentially affected by resonant interactions--- includes
collisions with the states
$\pi,K,\bar{K},\eta,N,\bar{N},\Delta,\bar{\Delta}$. A description of
their interactions has been developed in
Refs.~\cite{GarciaRecio:2008dp,Abreu:2011ic,Romanets:2012hm,Abreu:2012et,GarciaRecio:2012db,Garcia-Recio:2013gaa,Tolos:2013kva,Torres-Rincon:2014ffa,Tolos:2013gta}
using effective field theory. Moreover, after the application of an
effective theory, one should implement a unitarization method to the scattering amplitudes
 to better control the behavior of the cross sections at moderates energies.

The details of the interaction for the four heavy states follows
quite in parallel by virtue of the ``heavy-quark spin-flavor
symmetry''. It accounts for the fact that if the heavy masses are
much larger than any other typical scale in the system, like
$\Lambda_{QCD}$, temperature and the light hadron masses, then the
physics of the heavy subsystem is decoupled from the light sector,
and the former is not dependent on the mass nor on the spin of the
heavy particle. This symmetry is exact in the ideal limit $m_Q
\rightarrow \infty$, with $m_Q$ being the mass of the heavy quark
confined in the heavy hadron. In the opposite limit $m_Q \rightarrow
0$, one can exploit the chiral symmetry of the QCD Lagrangian to
develop an effective realization for the light particles. This
applies to the pseudoscalar meson octet ($\pi,K,\bar{K},\eta$).
Although both symmetries are broken in nature (as in our approach,
when implementing physical masses), the construction of the
effective field theories incorporates the breaking of these
symmetries in a controlled way. In particular, it provides a
systematic expansion in powers of $1/m_H$ (inverse heavy-meson mass)
and powers of $p, m_l$ (typical momentum and mass of the light
meson). Following these ideas, we use two effective Lagrangians for
the interaction of a heavy meson with light mesons and with baryons,
respectively.

In the scattering with light mesons, the scalar ($D$) and vector
($D^*$) mesons are much heavier than the pseudoscalar meson octet
($\pi,K,\bar{K},\eta$). The latter have, in addition, masses smaller
than the chiral scale $\Lambda_{\chi} \simeq 4 \pi f_\pi$, where
$f_\pi$ is the pion decay constant. In this case one can exploit
standard chiral perturbation theory for the dynamics of the (pseudo)
Goldstone bosons, and add the heavy-quark mass expansion up to the
desired order to account for the interactions with heavy mesons. In
our case the effective Lagrangian is kept to next-to-leading order
in the chiral expansion, but to leading order in the heavy-quark
expansion~\cite{Abreu:2011ic,Abreu:2012et}. From this effective
Lagrangian one can compute the tree-level amplitude (or potential),
which describes the scattering of a heavy meson off a light meson as
worked out in Refs.~\cite{Tolos:2013kva,Torres-Rincon:2014ffa}.

For the heavy meson--baryon interaction we use an effective
Lagrangian based on a low-energy realization of a $t-$channel vector
meson exchange between mesons and baryons. In the low-energy limit
the interaction provides a generalized Weinberg-Tomozawa contact
interaction as worked out in Refs.
~\cite{GarciaRecio:2008dp,Romanets:2012hm,GarciaRecio:2012db,Garcia-Recio:2013gaa}.
The effective Lagrangian obeys SU(6) spin-flavor symmetry in the
light sector, {plus heavy-quark spin symmetry} (HQSS) in the heavy sector (which is preserved
either the heavy quark is contained in the meson or in the baryon).

The tree-level amplitudes for meson-meson and meson-baryon
scattering have strong limitations in the energy range in which they
should be applied. It is limited for those processes in which the
typical momentum transfer is low, and below any possible resonance.
To increase the applicability of the tree-level scattering
amplitudes and restore exact unitarity for the scattering-matrix
elements, we apply a unitarization method, which consists in solving
a coupled-channel Bethe-Salpeter equation for the unitarized
scattering amplitude $T_{ij}$ using the potential $V_{ij}$ as a kernel,
\begin{equation}
 T_{ij} = V_{ij} + V_{ik} G_k T_{kj} \ ,
\end{equation}
where $G_k$ is the diagonal meson-meson (or meson-baryon) propagator
which is regularized by dimensional regularization in the
meson-meson (or meson-baryon) channel. We adopt the ``on-shell''
approximation to the kernel of the Bethe-Salpeter equation to reduce
it into a set of algebraic equations. We refer the reader to
Refs.~\cite{GarciaRecio:2008dp,Romanets:2012hm,GarciaRecio:2012db,Garcia-Recio:2013gaa,Tolos:2013kva,Torres-Rincon:2014ffa}
for technical details and individual results.

The unitarization procedure allows for the possibility of generating
resonant states as poles of the scattering amplitude $T_{ij}$ in the
complex plane. Even when these resonances are not explicit
degrees-of-freedom, and we do not propagate them in our PHSD
simulations, they are automatically incorporated into the two-body
interaction. This is an important extension, because such
(intermediate) resonant states will strongly affect the scattering
cross section of heavy mesons due to the presence of resonances,
subthreshold states (bound states), and other effects like the
opening of a new channel when a resonance is forming (Flatt\'e
effect).

The resulting (unitarized) cross sections for the binary scattering
of $D,D^*$ (with any possible charged states) with
$\pi,K,\bar{K},\eta,N,\bar{N},\Delta,\bar{\Delta}$ are implemented
in the PHSD code considering both elastic and inelastic channels.
About 200 different channels are taken into account. Although the
unitarization method helps to extend the validity of the tree-level
amplitudes into the resonant region, one cannot trust the final
cross sections for higher energies. Beyond the resonant region the
transition between the high and low energy regimes is
interpolated such that the cross sections are continuous.

\section{Dilepton production channels}
We recall that in the hadronic sector PHSD is equivalent to the
Hadron-String-Dynamics (HSD) transport approach \cite{HSD} that has
been used for the description of $pA$ and $AA$ collisions from SIS
to SPS energies and has lead to a fair reproduction of hadron
abundances, rapidity distributions and transverse momentum spectra
as well as dilepton spectra. In particular, HSD incorporates
off-shell dynamics for vector mesons and a set of vector-meson
spectral functions~\cite{Brat08dil} that covers possible scenarios
for their in-medium modification, i.e. in particular a collisional
broadening of the vector resonances. Note that in the off-shell
transport description, the hadron spectral functions change
dynamically during the propagation through the medium and evolve
towards the on-shell spectral function in the vacuum. The dilepton
production by a (baryonic or mesonic) resonance $R$ decay can be
schematically presented in the following way:
\begin{eqnarray}
 BB &\to&R X   \label{chBBR} \\
 mB &\to&R X \label{chmBR} \\
      && R \to  e^+e^- X, \label{chRd} \\
      && R \to  m X, \ m\to e^+e^- X, \label{chRMd} \\
      && R \to  R^\prime X, \ R^\prime \to e^+e^- X, \label{chRprd}
\end{eqnarray}
i.e. in a first step a resonance $R$ might be produced in
baryon-baryon ($BB$) or meson-baryon ($mB$) collisions.  Then this
resonance can couple to dileptons directly (\ref{chRd}) (e.g.,
Dalitz decay of the $\Delta$ resonance: $\Delta \to e^+e^-N$) or
decays to a meson $m$ (+ baryon) (\ref{chRMd}){, which} produce{s}
dileptons via direct decays ($\rho, \omega, \phi$) or Dalitz decays
($\pi^0, \eta, \omega$). The resonance $R$ might also decay into
another resonance $R^\prime$  (\ref{chRprd}) which later produces
dileptons via Dalitz decay.  Note, that in the combined model the
final particles -- which couple to dileptons -- can be produced also
via non-resonant mechanisms, i.e. `background' channels at low and
intermediate energies or string decay at high energies. In addition
to the hadronic channels above we account for the  '$4 \pi$' channels,
i.e. the dilepton production in the two-body reactions $\pi+\rho$,
$\pi+\omega$, $\rho+\rho$, $\pi+a_1$ as described in detail in Ref.
\cite{Olenasps}.
The latter provide the background from hadronic channels in the
intermediate mass regime 1.2 GeV $< M <$ 3 GeV \cite{Olenasps}, which is not
shown explicitly in this study since the contribution of '$4\pi$'
channels is much smaller than the contribution
from open charm decays and the QGP radiation.

We recall that the influence of in-medium effects on the
vector mesons ($\rho, \omega, \phi$) has been extensively studied
within the PHSD approach in the past (cf. Refs. \cite{Brat08dil,Olenasps,PHSDreview})
and it has been shown that the collisional broadening scenario for the
in-medium  vector-meson spectral functions is
consistent with  experimental dilepton data from SPS to LHC energies in line with the findings by other groups \cite{ChSym}. Accordingly,  in the present study we will adopt the
collisional broadening scenario for the vector-meson spectral functions
as the  'default' scenario.

In order to address the electromagnetic radiation of the partonic
phase, off-shell cross sections of $q\bar q\to\gamma^*$, $q\bar
q\to\gamma^*g$ and $qg\to\gamma^*q$ ($\bar q g\to\gamma^* \bar q$)
reactions taking into account the effective propagators for quarks
and gluons from the DQPM have been calculated in
Ref.~\cite{olena2010}. Here $\gamma ^*$ stands for the $e^+e^-$ or
$\mu^+ \mu^-$ pair. Dilepton production in the QGP - as created in
early stages of heavy-ion collisions - is calculated by implementing
these off-shell processes into the PHSD transport approach on the
basis of the same partonic propagators as used for the
time-evolution of the partonic system. For a review on
electromagnetic production channels within PHSD we refer the reader
to Ref. \cite{PHSDreview} and for the details of the dilepton cross sections
from off-shell partonic channels to  Appendix A.

\section{Results for heavy-ion reactions}\label{results}

So far we have described the interactions of the heavy flavor -
produced in relativistic heavy-ion collisions - with partonic and
hadronic degrees-of-freedom. Since the  matter produced in heavy-ion
collisions is extremely dense, the interactions with the bulk matter
suppresses heavy flavors at high-$\rm p_T$. On the other hand, the
partonic or nuclear matter is accelerated outward (exploding), and a
strong flow is generated via the interactions of the bulk particles
and the repulsive scalar interaction for partons. Since the heavy
flavor strongly interacts with the expanding matter, it is also
accelerated outwards. Such effects of the medium on the heavy-flavor
dynamics are expressed in terms of the nuclear modification factor
defined as
\begin{eqnarray}
R_{\rm AA}(\rm p_T)\equiv\frac{dN_{\rm AA}/d{\rm p_T}}{N_{\rm binary}^{\rm AA}\times dN_{\rm pp}/d{\rm p_T}},
\label{raa}
\end{eqnarray}
where $N_{\rm AA}$ and $N_{\rm pp}$ are, respectively, the number of
particles produced in heavy-ion collisions and that in p+p
collisions, and $N_{\rm binary}^{\rm AA}$ is the number of binary
nucleon-nucleon collisions in the heavy-ion collision for the
centrality class considered. Note that if the heavy flavor does not
interact with the medium in heavy-ion collisions, the numerator of
Eq.~(\ref{raa}) will be similar to the denominator. For the same
reason, a $\rm R_{\rm AA}$ smaller (larger) than one in a specific
$\rm p_T$ region implies that the nuclear matter suppresses
(enhances) the production of heavy flavors in that transverse
momentum region.

In noncentral heavy-ion collisions the produced  matter expands
anisotropically due to the different pressure gradients between in
plane and out-of plane. If the heavy flavor interacts strongly with
the nuclear matter, then it also follows this anisotropic motion to
some extend. The anisotropic flow is expressed in terms of the
elliptic flow $v_2$ which reads
\begin{eqnarray}
v_2({\rm p_T})\equiv\frac{\int d\phi \cos2\phi (dN_{\rm AA}/d{\rm p_T}d\phi)}{2\pi dN_{\rm AA}/d{\rm p_T}},
\end{eqnarray}
where $\phi$ is the azimuthal angle of a particle in momentum space.

\subsection{Nuclear modification of dielectrons from heavy flavor}

In this subsection we focus on the $c{\bar c}$ dynamics and the dielectrons produced from heavy flavor pairs and their modification in relativistic heavy-ion collisions.

\begin{figure} [h!]
\centerline{
\includegraphics[width=8.6 cm]{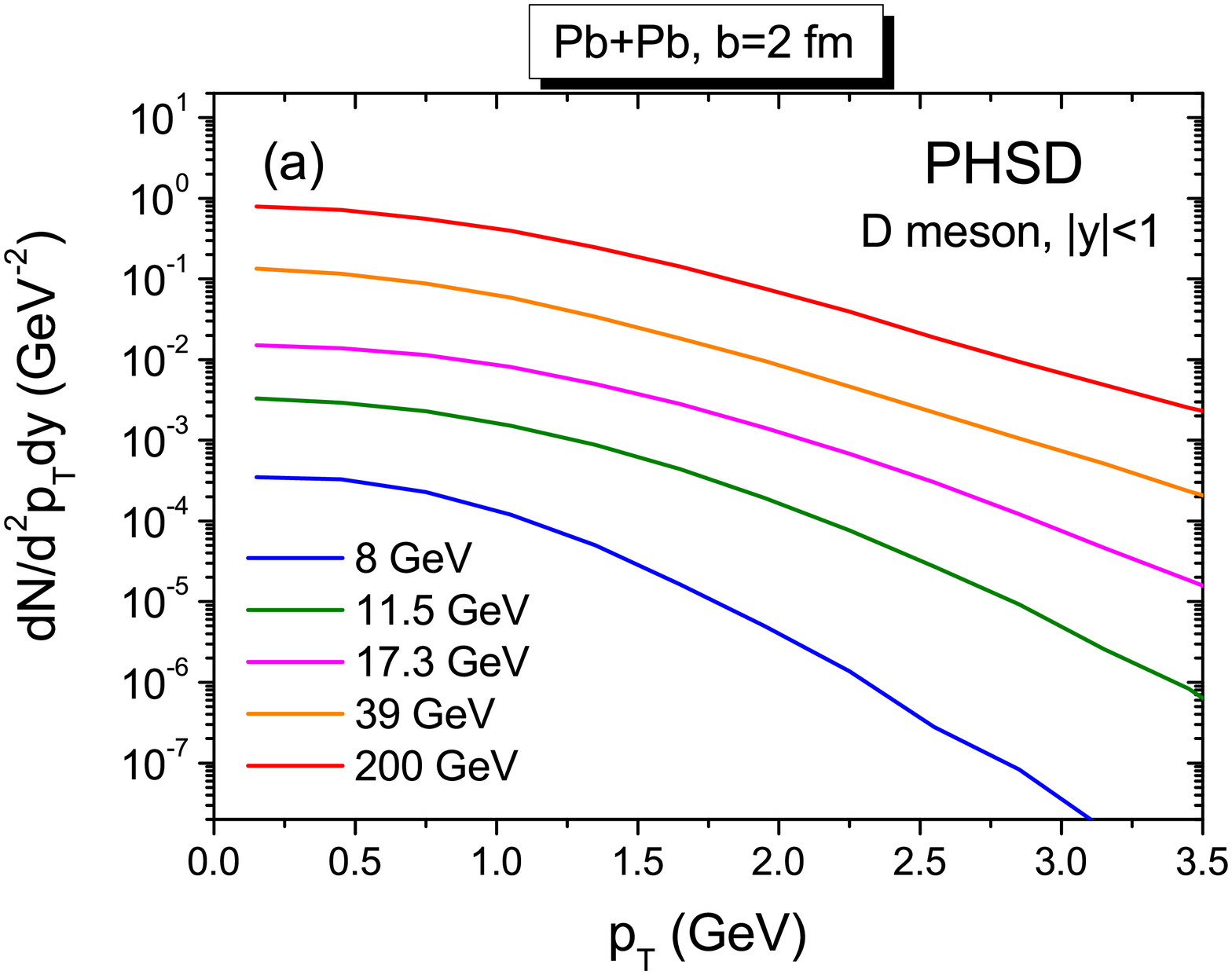}}
\centerline{
\includegraphics[width=8.6 cm]{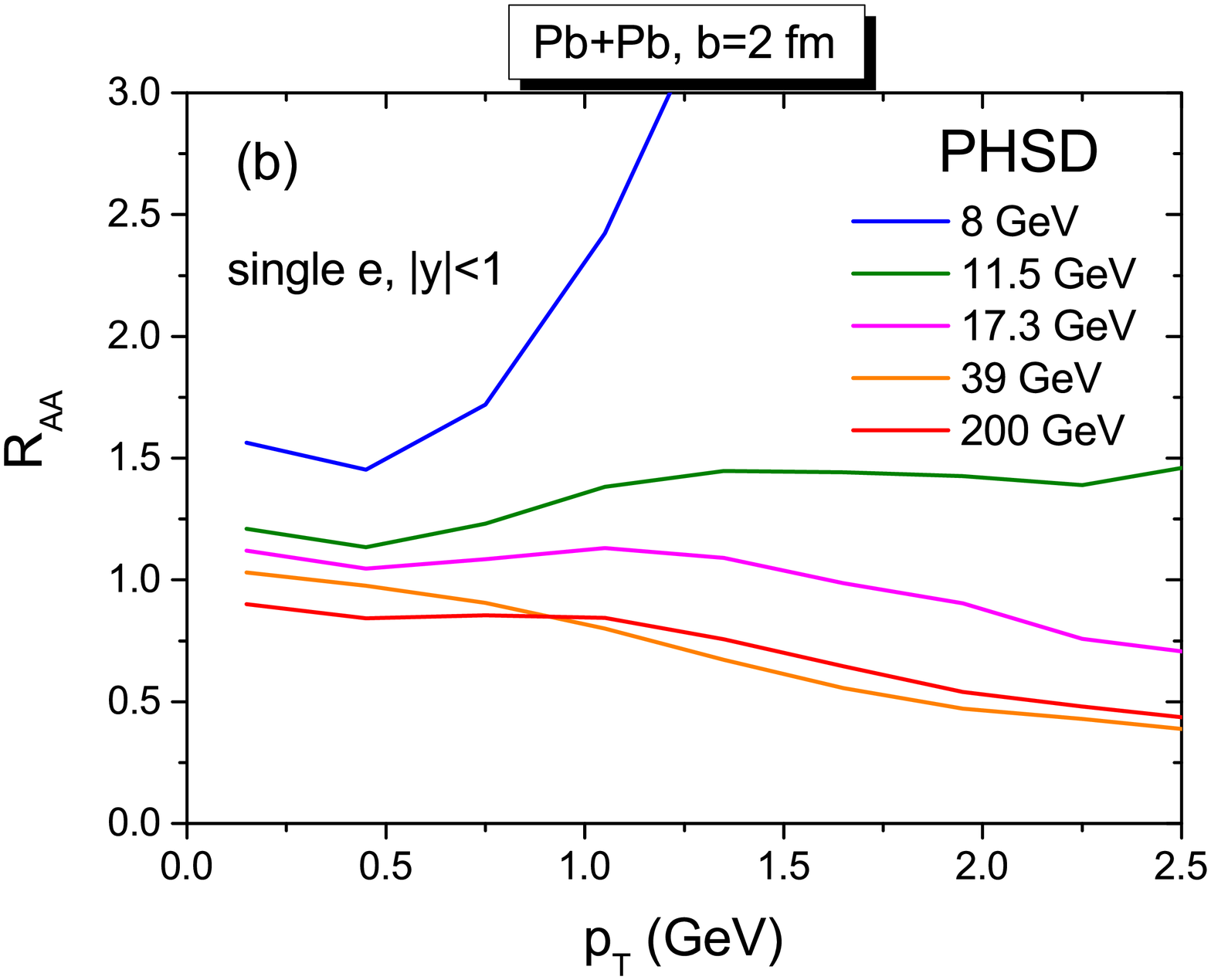}}
\caption{The transverse momentum spectra of $D$ mesons (a) and the  $R_{AA}(p_T)$ of  single electrons from semi-leptonic decay of $D$ mesons (b) as a function of the transverse momentum $p_T$ in central Pb+Pb collisions from PHSD at  $\sqrt{s_{\rm NN}}$ = 8, 11.5, 17.3, 39 and 200 GeV at midrapidity.} \label{fig2}
\end{figure}

Fig. \ref{fig2} (a) shows the transverse momentum spectra of $D$ mesons in central Pb+Pb collisions at $\sqrt{s_{\rm NN}}$ = 8, 11.5, 17.3, 39, and 200 GeV for $|y| < $ 1.
Since the cross section for charm production increases with collision energy as shown in Fig. \ref{fig1} (a), the transverse momentum spectrum of $D$ meson enhances strongly with increasing collision energy and also becomes harder.

Fig. \ref{fig2} (b) {displays} the nuclear modification factor of single electrons from $D$ meson semi-leptonic decays at mid-rapidity ($|y|<1$) for the same set of central Pb+Pb collisions.
 We mention that for the semi-leptonic decays of heavy flavors we use the subroutine `pydecay' of the PYTHIA event generator~\cite{Sjostrand:2006za}. Contrary to the $R_{\rm AA}$ at RHIC and LHC energies we find ratios well above unity at $\sqrt{s_{\rm NN}}$ = 8 and 11.5 GeV  which implies an enhancement of the yield (at higher momenta) rather than the familiar suppression at RHIC and LHC.
The enhanced $R_{\rm AA}$ at low energies (8 and 11.5 GeV) may be dominantly attributed to the Fermi motion of nucleons in the colliding nuclei, which does not exist in p+p collisions and slightly increases the collision energy in binary nucleon-nucleon scattering.
Since the collision energies are close to the threshold energy for charm-pair production, where the production cross section increases rapidity as shown in Fig.~\ref{fig1} (a), a small enhancement of the collision energy gives a sizeable increase of the charm production and subsequently the decay products.
We note in passing that the $R_{\rm AA}$ of single electrons at $\sqrt{s_{\rm NN}}$ = 39 and 200 GeV is consistent with our recent results in Ref.~\cite{Song2016}, where the $R_{\rm AA}$ is shown also for higher transverse momenta.

\begin{figure}[h!]
\centerline{
\includegraphics[width=8.6 cm]{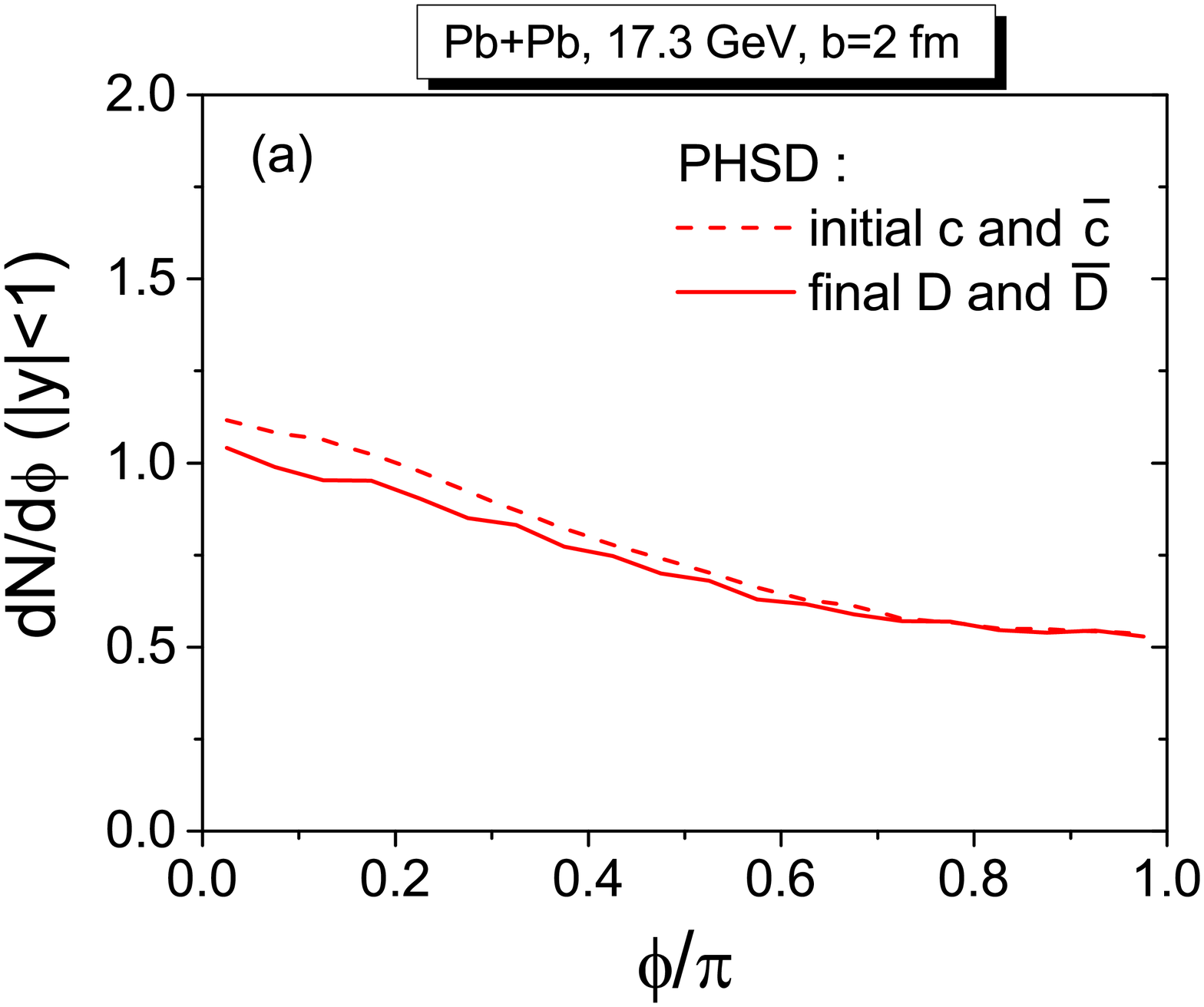}}
\centerline{
\includegraphics[width=8.6 cm]{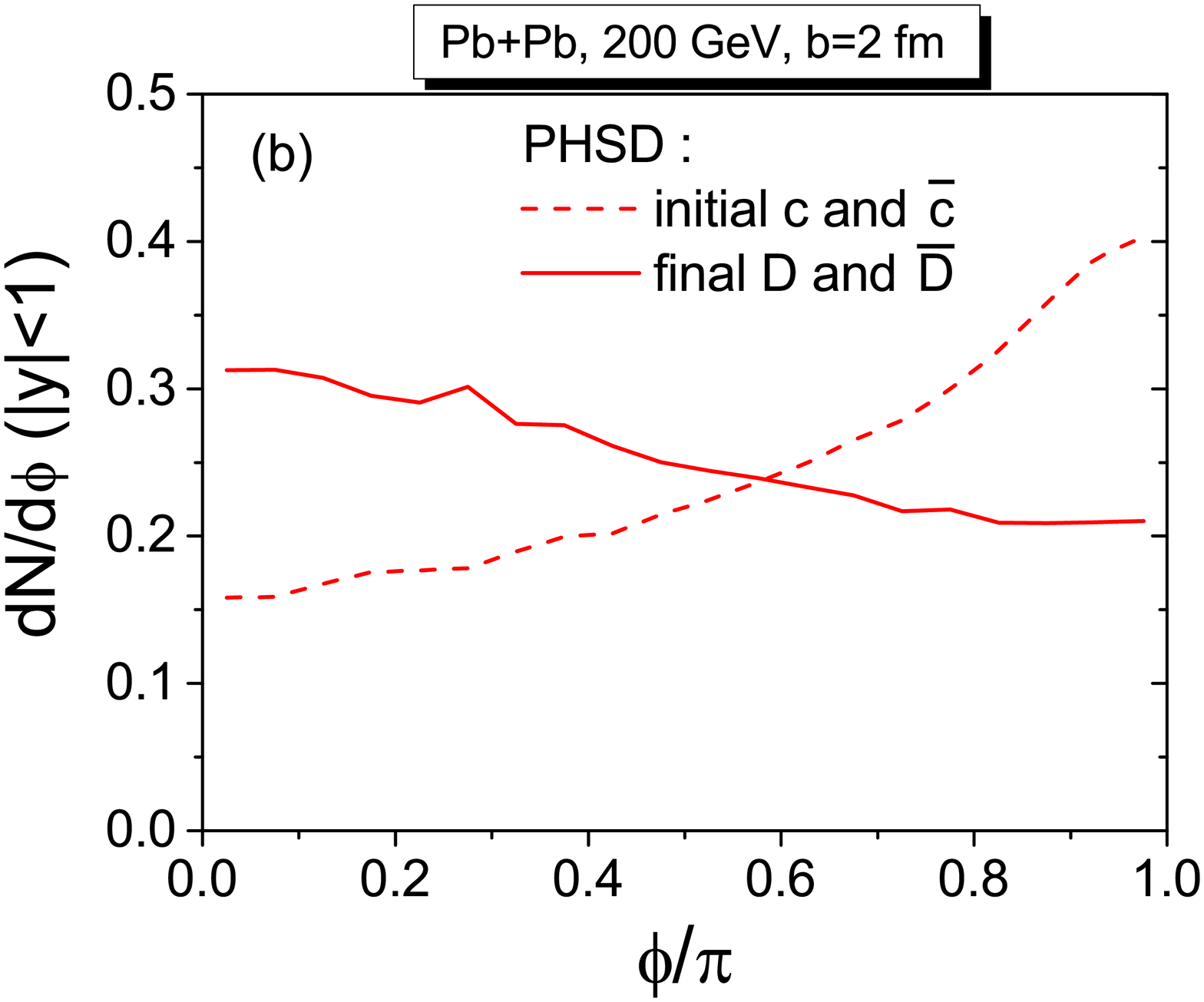}}
\caption{Azimuthal angular distribution between the transverse
momentum of a heavy-flavor meson and that of an antiheavy-flavor
meson for each heavy flavor pair {at midrapidity $(|y|<1)$} before (dashed lines) and after the interactions
with the medium (solid lines) in {central Pb+Pb collisions at $\sqrt{s_{\rm NN}}$ = 17.3 (a) and 200 GeV} (b).  }
\label{fig3}
\end{figure}

Since heavy flavor is always produced by pairs, there is an angular correlation between the heavy quark and heavy antiquark. If the heavy quark and  antiquark from the same pair (through semi-leptonic decays) produce a positron and an electron, respectively, the produced dielectron also has an angular correlation.
On the other hand, the  matter produced in heavy-ion collisions  changes the transverse momentum of each heavy flavor and consequently  also the angular correlation of the heavy flavor pair.
It has been suggested  that the analysis of the azimuthal
angular correlation might provide information on the energy loss
mechanism of heavy quarks in the QGP~\cite{Cao:2015cba}, because
stronger interactions should result in less pronounced angular
correlations. Since in the PHSD we can follow up the fate of an
initial heavy quark-antiquark pair throughout the partonic
scatterings, the hadronization and final hadronic rescatterings, the
microscopic calculations allow to shed some light on the correlation
between the in-medium interactions and the final angular
correlations.

Fig. \ref{fig3} shows the azimuthal angular distribution between the
transverse momentum of {charm ($D$)} and that of {anti-charm ($\bar{D}$)} for each {charm} pair {at midrapidity $(|y|<1)$} before (dashed lines) and after
the interactions with the medium (solid lines) in central Pb+Pb
collisions {at $\sqrt{s_{\rm NN}}$ = 17.3 and 200 GeV}.
The azimuthal angle between the initial charm and anti-charm quarks is provided by the PYTHIA event generator and
peaks around $\phi= 0$ for $\sqrt{s_{\rm NN}}$ = 17.3 GeV, while we find a maximum around $\phi= \pi$  for $\sqrt{s_{\rm NN}}$ = 200 GeV.
After the interaction with the hadronic and partonic matter, however, the azimuthal angle between the $D$ and $\bar{D}$ has a maximum  near $\phi= 0$ at both collision energies.
In other words, the azimuthal angle  changes little in low-energy collisions, but considerably in high-energy collisions. As shown in our previous study~\cite{Song:2015ykw} the shift of the maximum in the azimuthal angle from $\pi$ to $0$ at $\sqrt{s_{\rm NN}}=$ 200 GeV can be attributed to the strong interaction of charm with radial flow.

\begin{figure}[h!]
\centerline{
\includegraphics[width=8.6 cm]{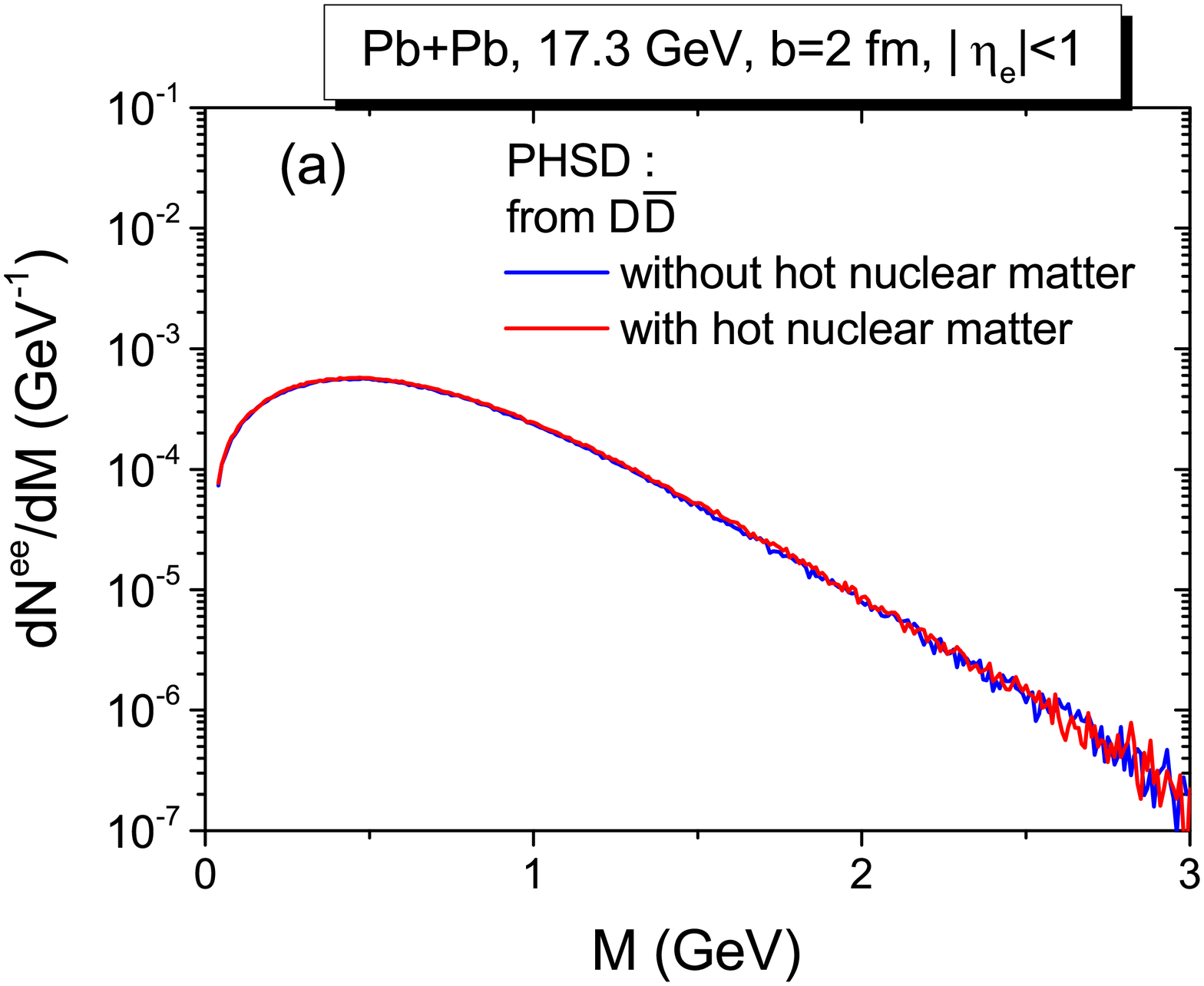}}
\centerline{
\includegraphics[width=8.6 cm]{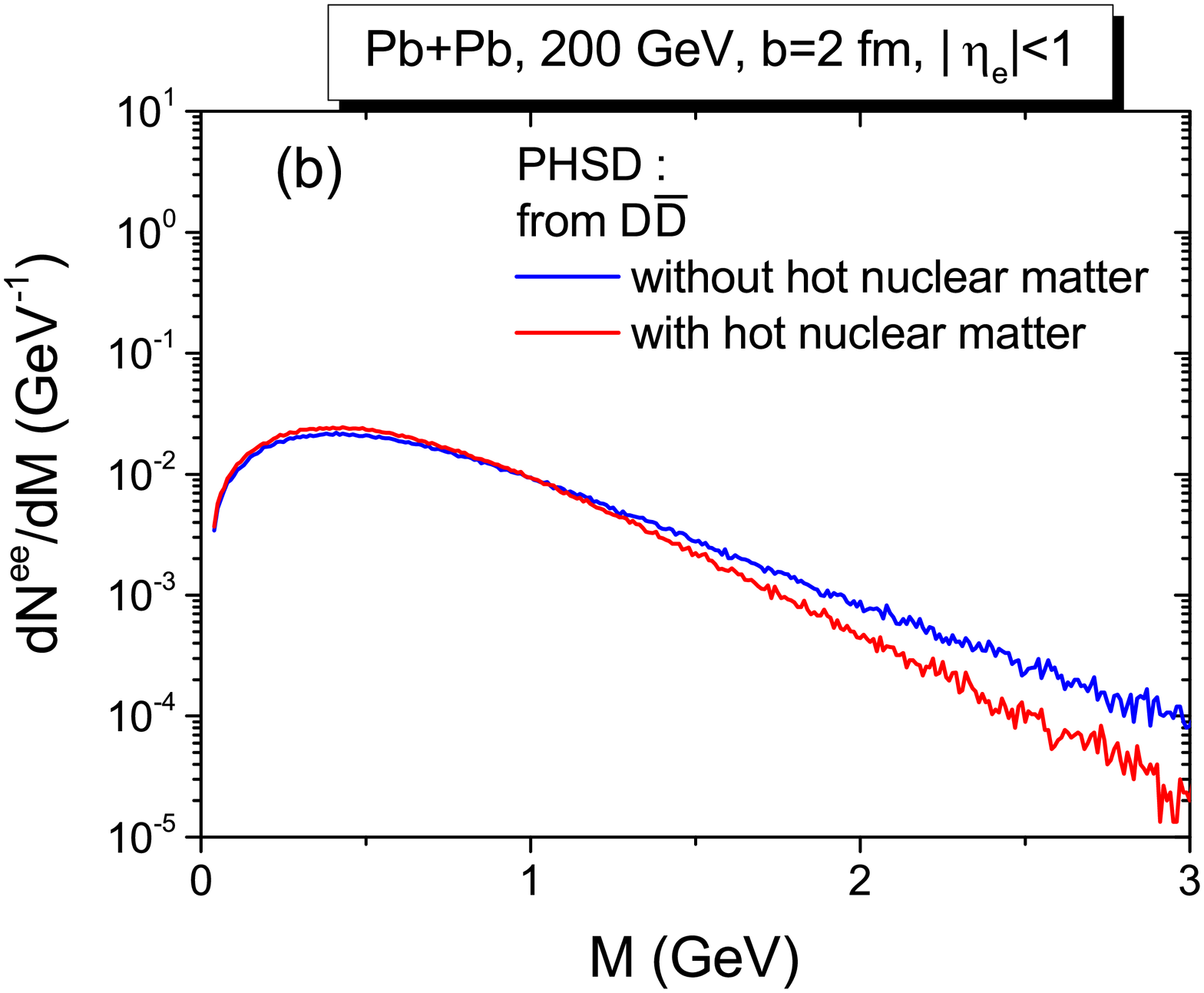}}
\caption{Invariant mass spectra of dielectrons from charm pairs with (red lines) and without  the interactions with the hot medium (blue lines) in central Pb+Pb collisions at $\sqrt{s_{\rm NN}}$ = 17.3 (a) and 200 GeV (b).}
\label{fig4}
\end{figure}

Fig.~\ref{fig4} shows the invariant mass spectra of dielectrons from charm pairs with (red lines) and without the interactions with hot medium (blue lines) in central Pb+Pb collisions at $\sqrt{s_{\rm NN}}$ = 17.3 (a) and 200 GeV (b).
We can see that the invariant mass spectrum of dielectrons changes little for $\sqrt{s_{\rm NN}}$ = 17.3 GeV, while it is considerably suppressed at large invariant mass at $\sqrt{s_{\rm NN}}$ = 200 GeV.
This suppression can be understood from Figs.~\ref{fig2} and \ref{fig3}, considering that
the invariant mass of the dielectron depends on the momenta of electron and positron and also on the angle between them.
Figs.~\ref{fig2} and \ref{fig3} clearly show that the momenta of electron and positron are suppressed and the azimuthal angle between them decreases at $\sqrt{s_{\rm NN}}$ = 200 GeV;  both effects decrease the invariant mass of the dielectron.
On the other hand, the momenta of electron and positron and the azimuthal angle do not change much at $\sqrt{s_{\rm NN}}$ = 17.3 GeV such that the dielectron spectrum stays approximately unchanged.

\subsection{Excitation function of dielectron production in Pb+Pb collisions from $\sqrt{s_{\rm NN}}=$8 to 200 GeV}

As mentioned in the previous sections, the dileptons produced in relativistic heavy-ion collisions can be  classified into three parts: i) dileptons from heavy flavor pairs, ii) from partonic scatterings in the QGP phase, and iii)  from hadronic interactions in the hadronic (HG) phase.
In this subsection we compare the separate contributions in central Pb+Pb collisions at various energies from {8} to 200 GeV.

\begin{figure} [h!]
\centerline{
\includegraphics[width=8.6 cm]{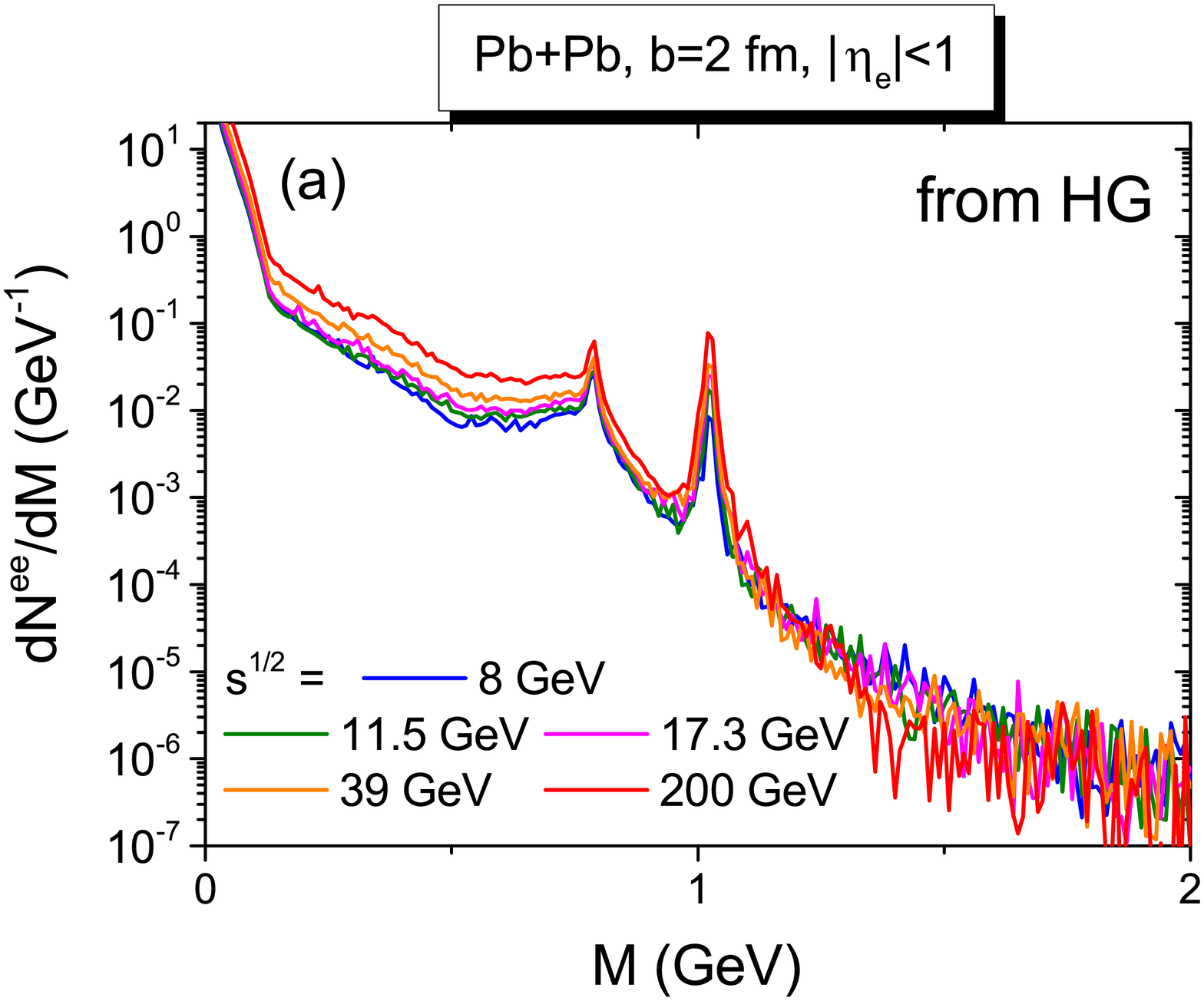}}
\centerline{
\includegraphics[width=8.6 cm]{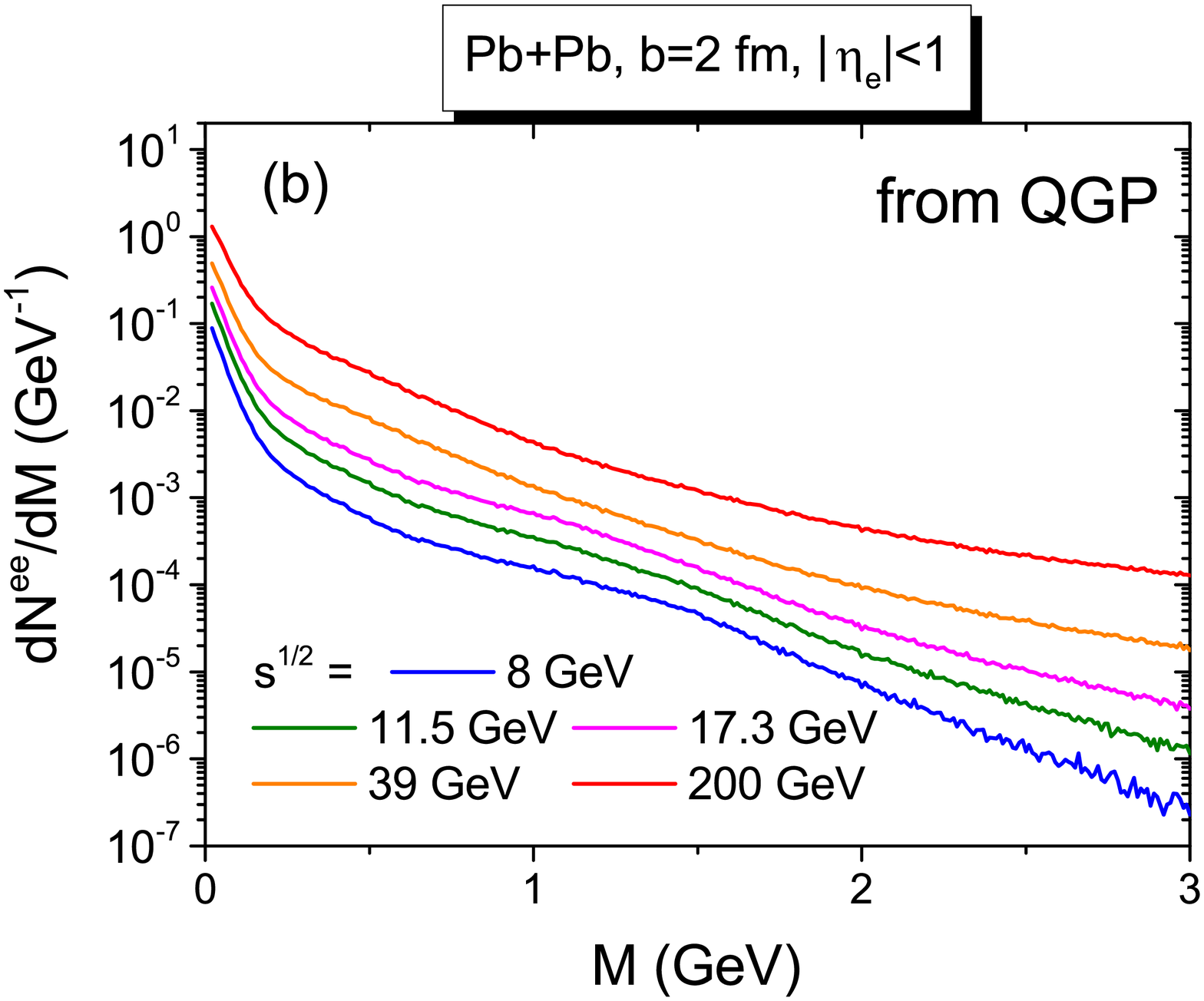}}
\centerline{
\includegraphics[width=8.6 cm]{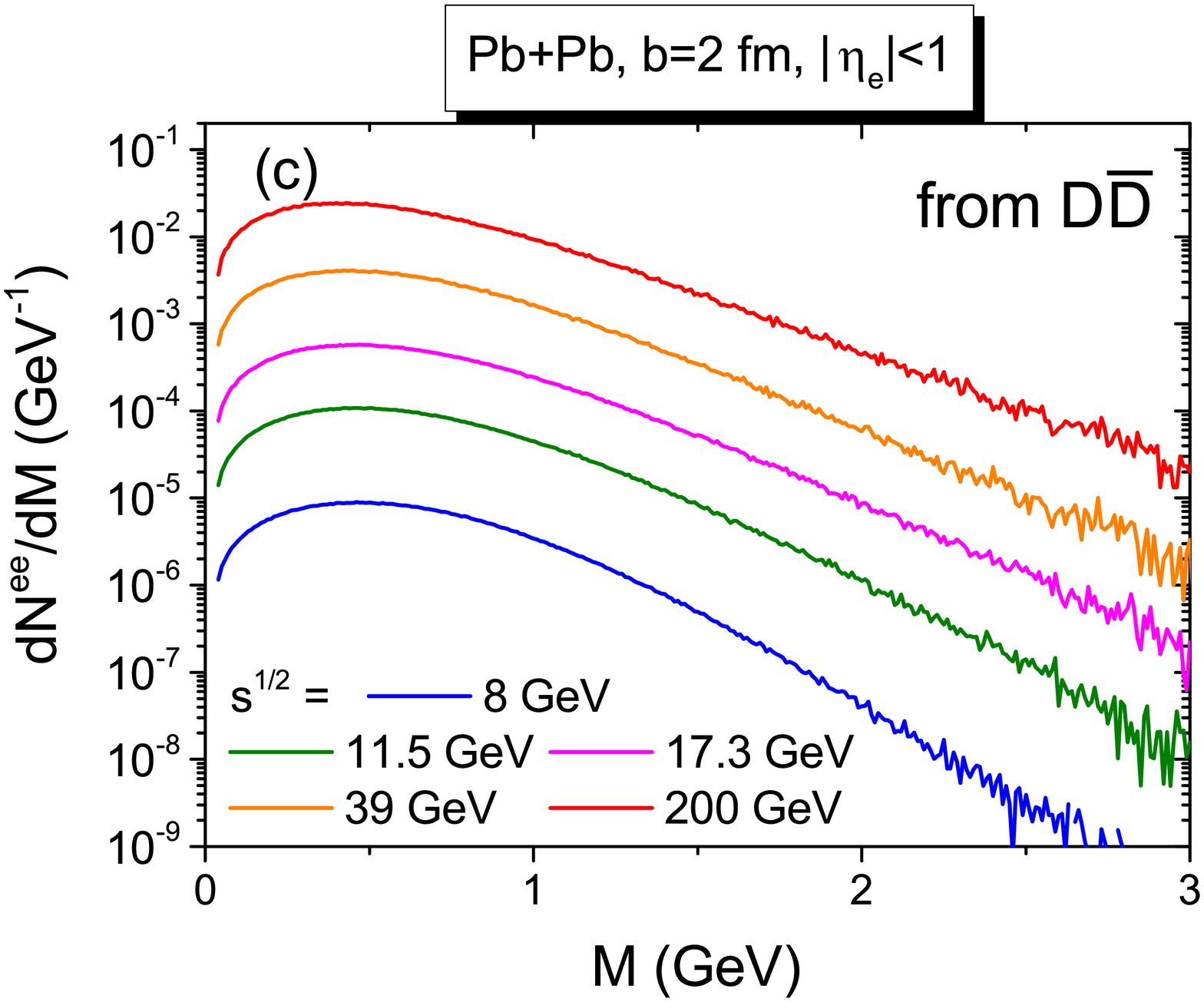}}
\caption{The invariant mass spectra of dileptons from the hadronic sources (HG) (a), the QGP (b), and $D\bar{D}$ pairs (c) in central Pb+Pb collisions at $\sqrt{s_{\rm NN}}$ = 8, 11.5, 17.3, 39 and 200 GeV from the PHSD.} \label{fig5}
\end{figure}

Fig.~\ref{fig5} shows the dielectron mass spectra from hadronic channels (a), from partonic interactions in the QGP (b), and from the semi-leptonic decays of $D\bar{D}$ pairs (c) in central Pb+Pb collisions at $\sqrt{s_{\rm NN}}$ = 8, 11.5, 17.3, 39, and 200 GeV at mid-pseudorapidity $|\eta_e|<1$ for the leptons.
We find that the contribution from the hadronic channels increases moderately with collision energy (in line with the hadron abundances), the contribution from the QGP raises more steeply (in line with the enhanced space-time volume of the QGP phase) and that from $D\bar{D}$ pairs is most dramatically increasing (in line with the number of $c{\bar c}$ pairs, cf. Fig. 1b)). Accordingly,  the contribution from heavy flavor is small at low-energy collisions, but becomes more and more important with increasing collision energy in competition with the production from the QGP channels.

\begin{figure*}[h!]
\centerline{
\includegraphics[width=8.6 cm]{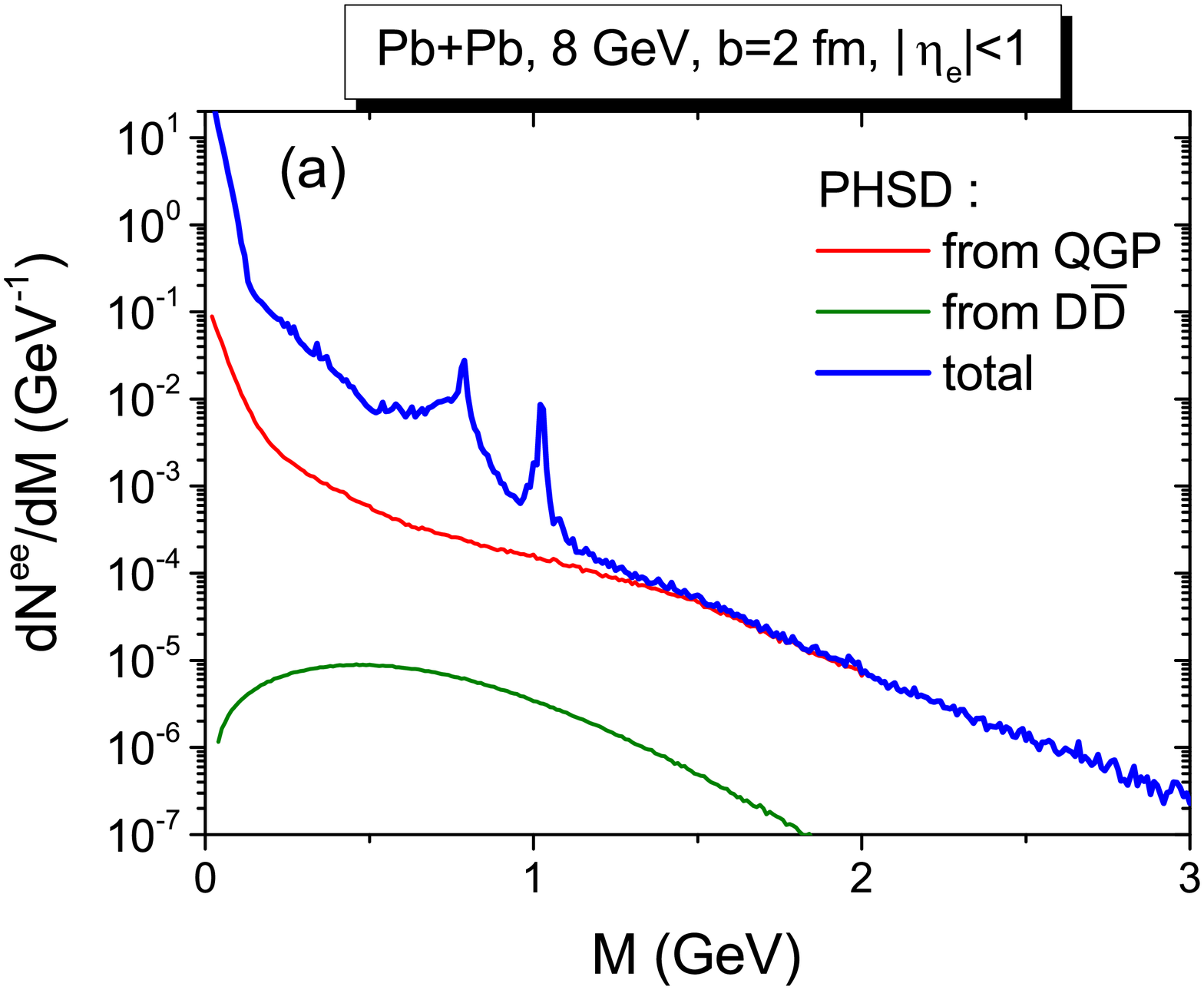}
\includegraphics[width=8.6 cm]{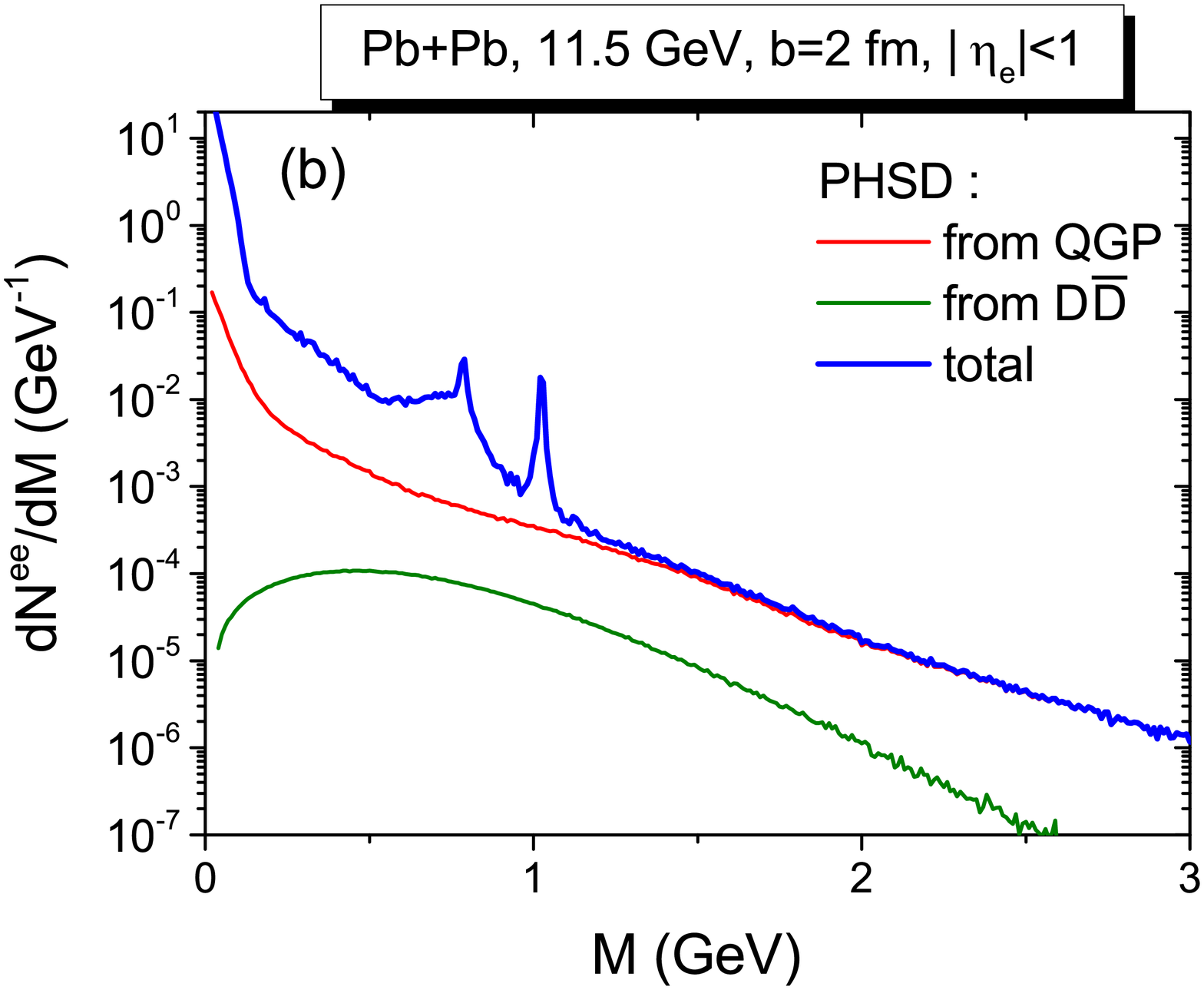}}
\centerline{
\includegraphics[width=8.6 cm]{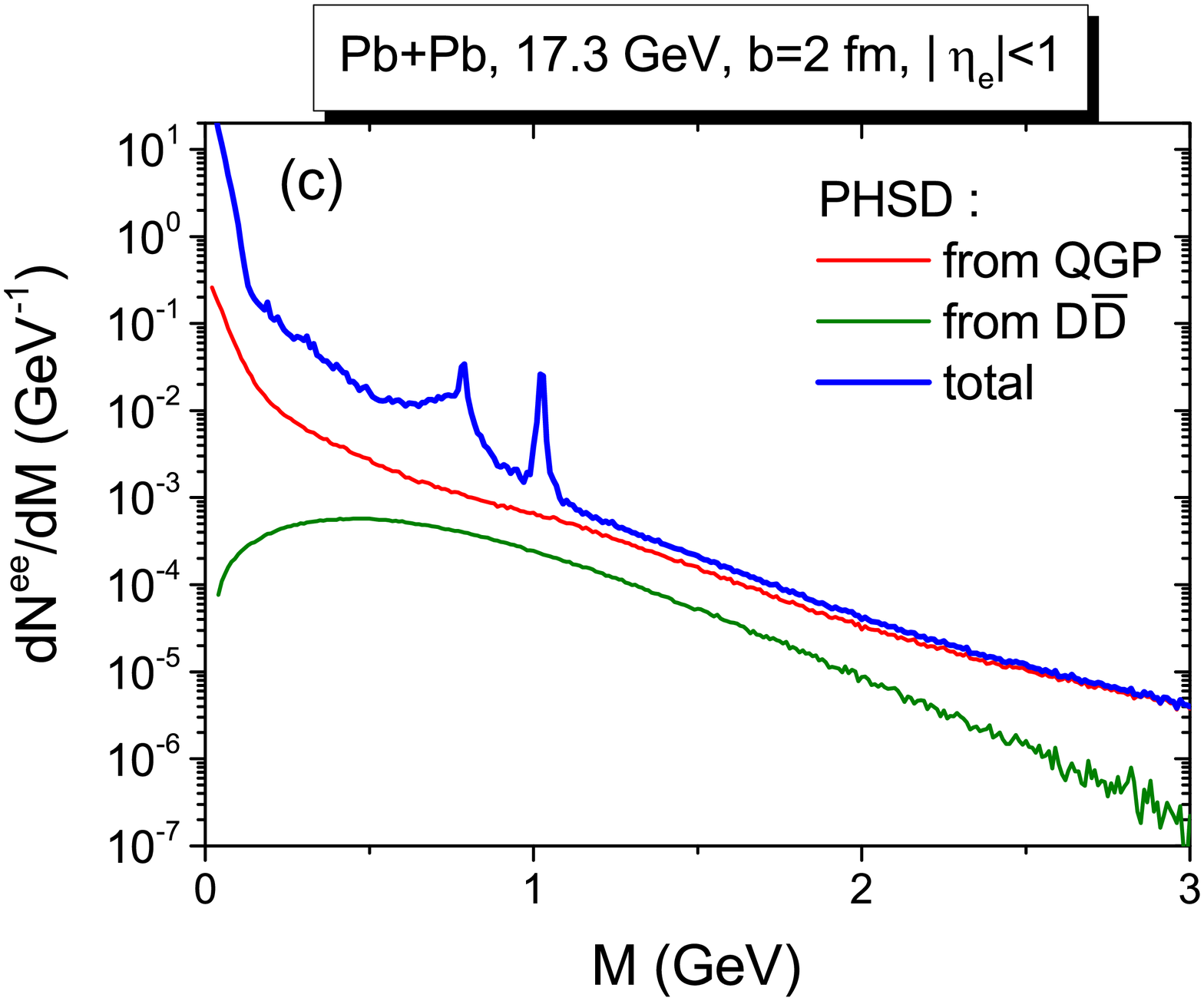}
\includegraphics[width=8.6 cm]{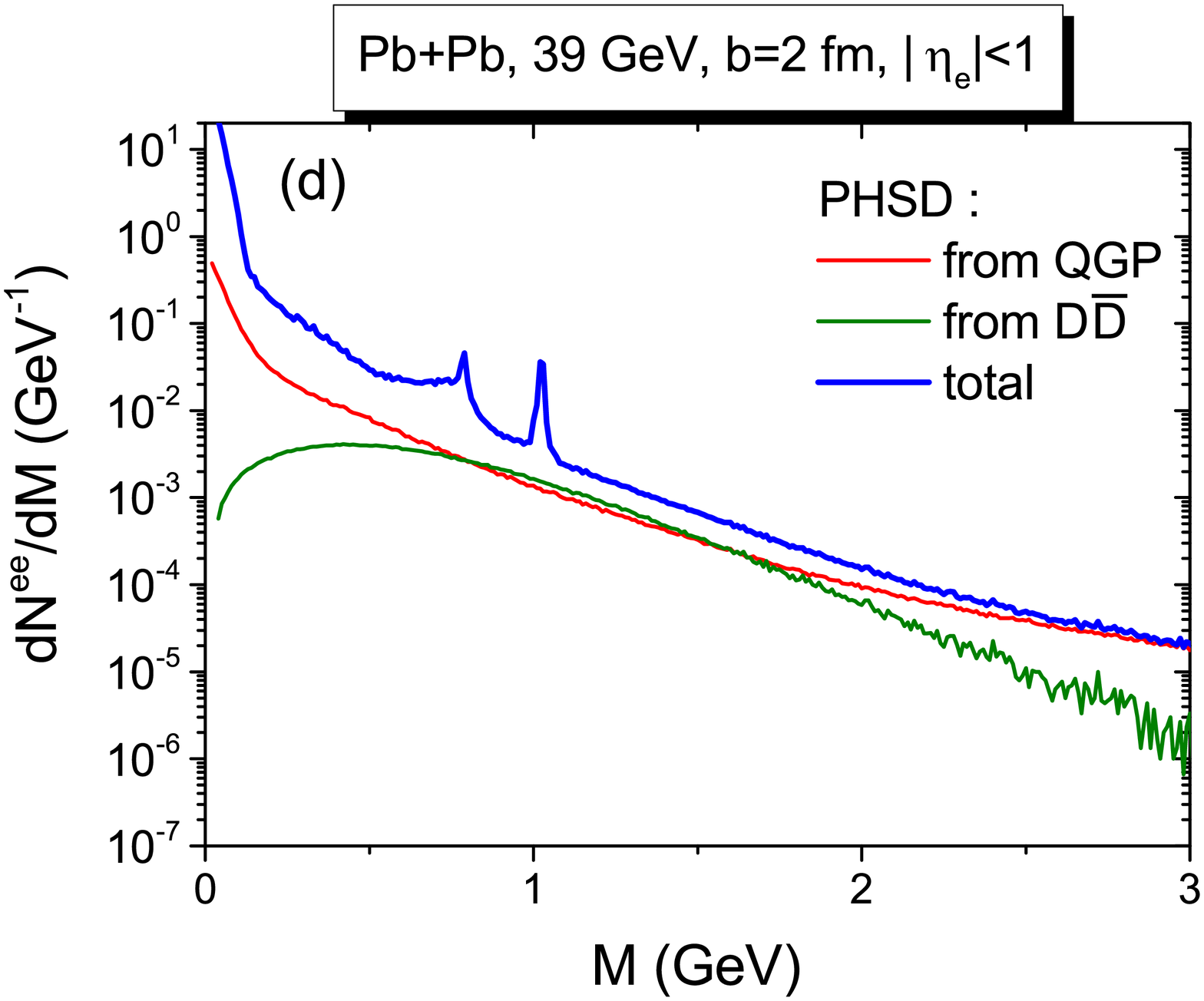}}
\centerline{
\includegraphics[width=8.6 cm]{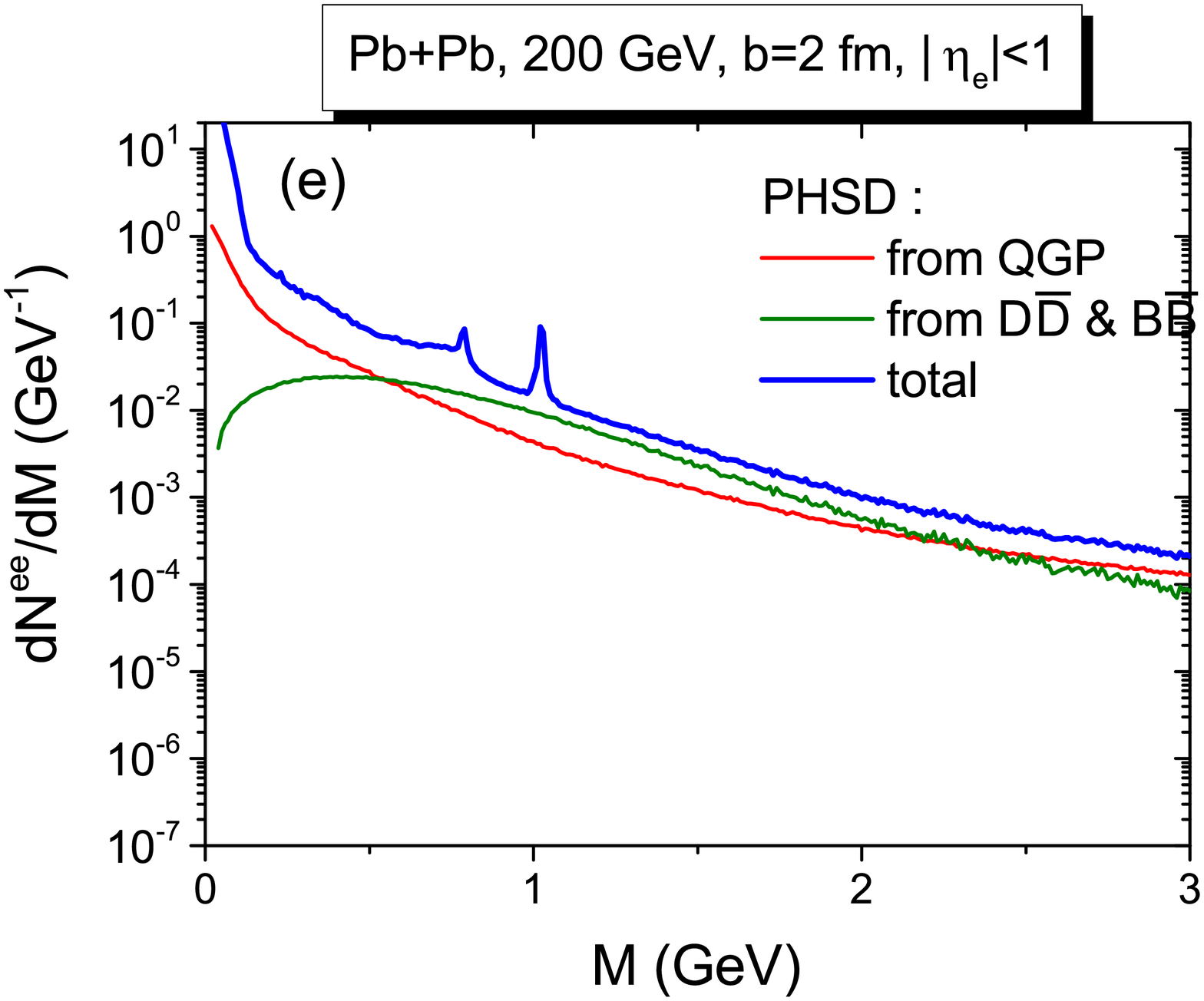}}
\caption{The invariant mass spectra of dileptons from partonic interactions (red lines) and from $D\bar{D}$ pairs (green lines) together with total dielectron spectrum including hadronic contributions (blue lines) in central Pb+Pb collisions at $\sqrt{s_{\rm NN}}$ = 8, 11.5, 17.3, 39 and 200 GeV from the PHSD at mid-pseudorapidity for the leptons.}
\label{fig6}
\end{figure*}

In order to show the separate contributions explicitly, we compare in Fig.~\ref{fig6} the contributions from the QGP (red lines) and from $D\bar{D}$ pairs (green lines)  with the total dielectron spectrum (blue lines) at different collision energies for central Pb+Pb collisions.
In low-energy collisions the dielectrons from hadronic channels dominate in the low-mass region and those from partonic interactions dominate in the intermediate-mass range while the contribution from $D\bar{D}$ pairs is negligible.
With increasing collision energy  the contribution from $D\bar{D}$ pairs becomes more and more significant and comparable to that from partonic interactions at $\sqrt{s_{\rm NN}} \approx$  39 GeV in the intermediate-mass range.
Finally, it overshines the partonic contribution at $\sqrt{s_{\rm NN}}$ = 200 GeV (and above).

\begin{figure} [h]
\centerline{
\includegraphics[width=8.6 cm]{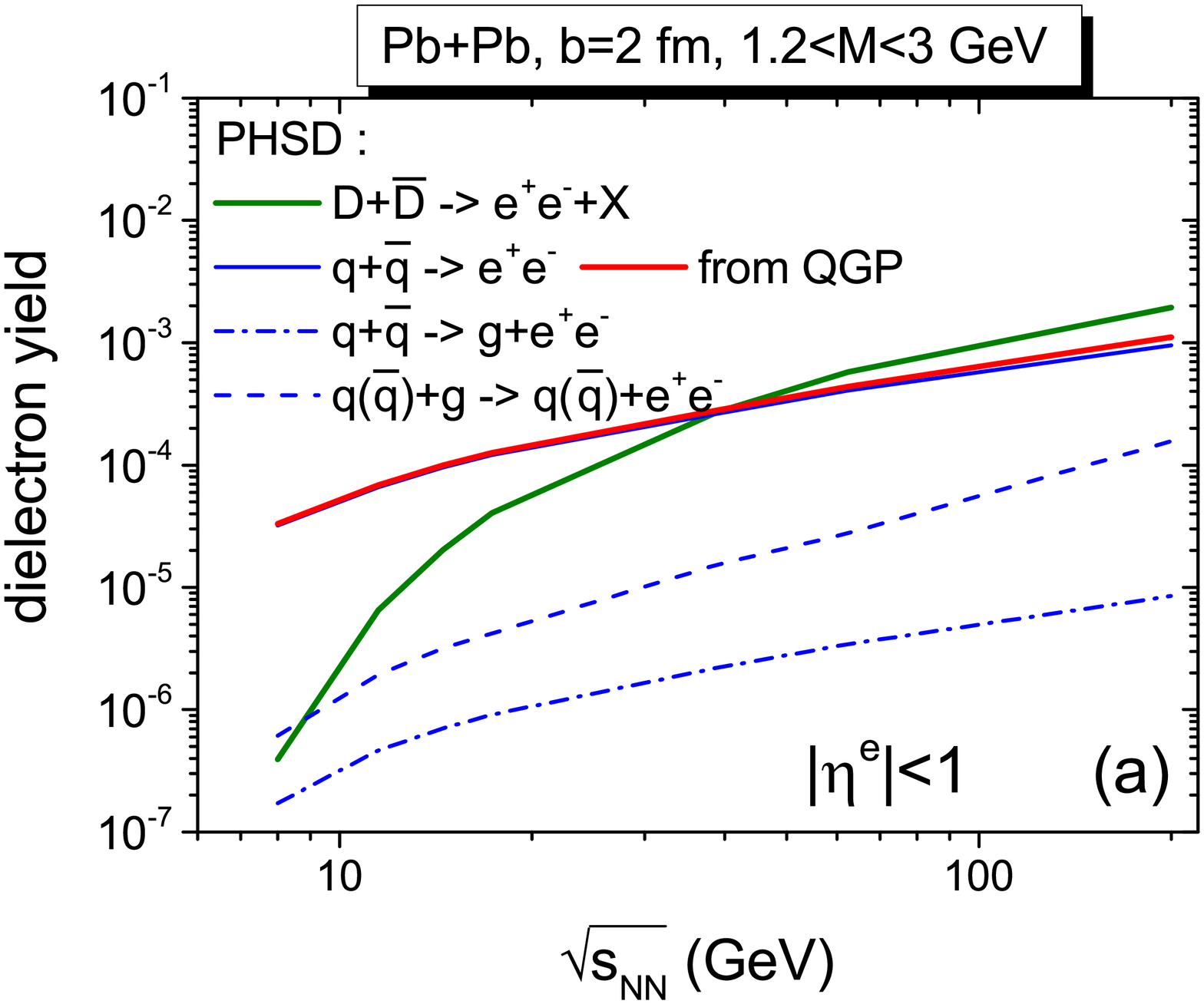}}
\centerline{
\includegraphics[width=8.6 cm]{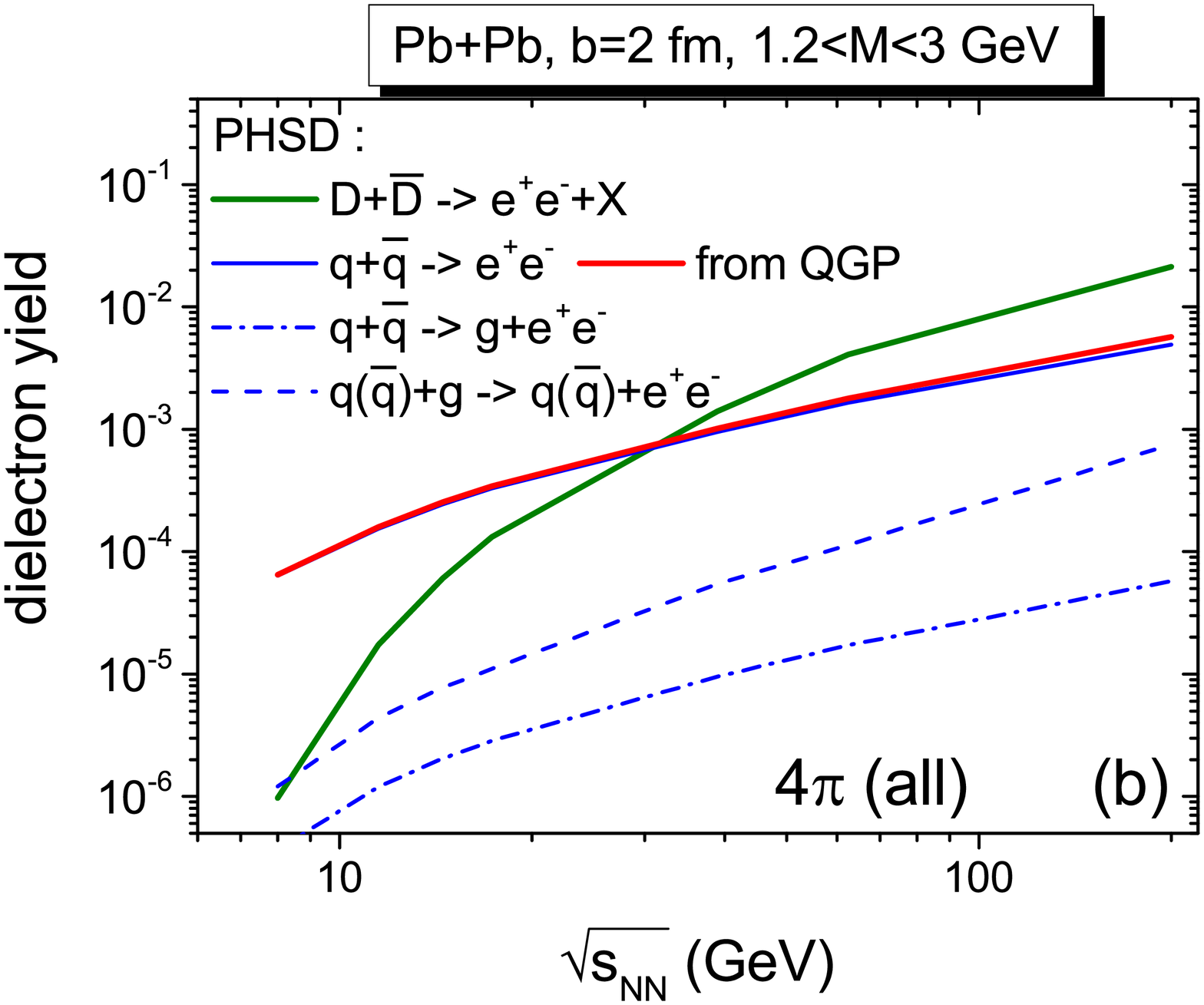}}
\caption{The contributions to intermediate-mass dielectrons (1.2 GeV $< M <$ 3 GeV) from $D\bar{D}$ pairs (green lines), different channels of partonic interactions, $q+\bar{q}\rightarrow e^++e^-$, $q+\bar{q}\rightarrow g+e^++e^-$, $q(\bar{q})+g\rightarrow q(\bar{q})+e^++e^-$ (see legend) as a function of $\sqrt{s_{\rm NN}}$ for Pb+Pb collisions at b=2 fm (for midrapidity leptons). The red solid line displays the sum of the partonic contributions.} \label{fig7}
\end{figure}

Fig.~\ref{fig7} compares the contributions from $D\bar{D}$ pairs (green lines) to three partonic channels, i.e. $q+\bar{q}\rightarrow e^++e^-$, $q+\bar{q}\rightarrow g+e^++e^-$, and $q(\bar{q})+g\rightarrow q(\bar{q})+e^++e^-$, for intermediate mass dileptons (1.2 GeV $< M <$ 3 GeV)  as a function of collision energy $\sqrt{s_{\rm NN}}$ for Pb+Pb collisions at b=2 fm. The figure clearly shows that the contribution from partonic interactions, especially from $q+\bar{q}\rightarrow e^++e^-$, dominates the intermediate-mass range in low-energy collisions.
However, the contribution from $D\bar{D}$ pairs rapidly increases with increasing collision energy, because the scattering cross section for charm production grows fast above the threshold energy as shown in Fig.~\ref{fig1} (a).
It overshines the contribution from partonic interactions around $\sqrt{s_{\rm NN}} \approx$ 40 GeV and dominates at higher energies.
Since the detectors of different collaborations have a different acceptance, we show in Fig.~\ref{fig7} (b) the results without any acceptance cuts, while Fig.~\ref{fig7} (a) shows the results for a mid-pseudorapidity cut on leptons of $|\eta^e|<1$.
However, the contributions from the partonic interactions and from $D\bar{D}$ pairs show a similar behavior in both cases.

One of most important issues in heavy-ion physics is to find and study the properties of partonic nuclear matter which is created in a small space-time volume in relativistic heavy-ion collisions. To this end one needs observables that are not blurred by hadronic interactions.
Our results in Figs.~\ref{fig6} and \ref{fig7} clearly demonstrate that the window to study partonic matter by dielectrons  at intermediate masses without substantial background from  heavy flavor decays opens for collision energies $\sqrt{s_{\rm NN}} <$ 40 GeV.

\subsection{Transverse mass spectra at midrapidity}

\begin{figure} [t!]
\centerline{
\includegraphics[width=8.6 cm]{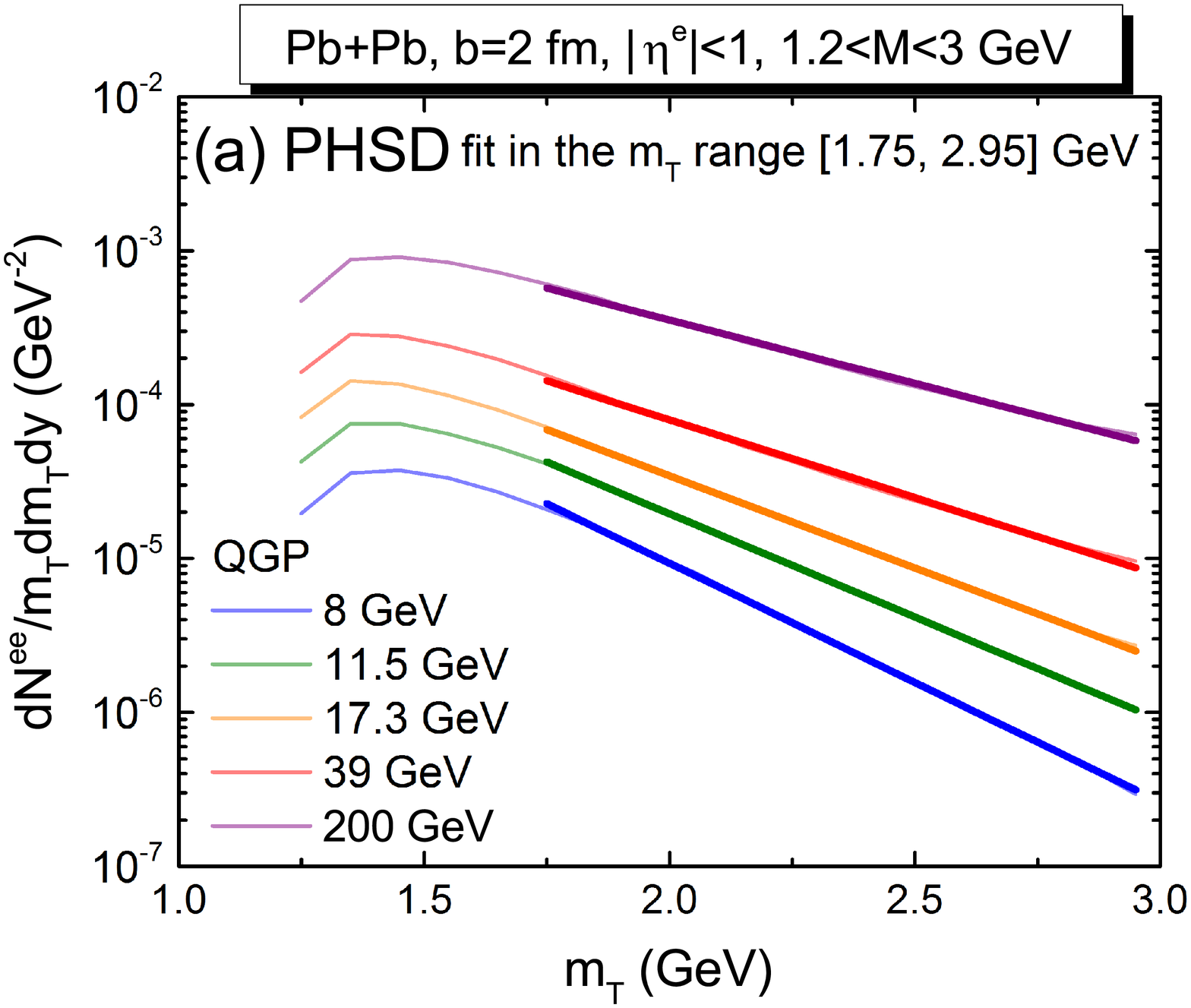}}
\centerline{
\includegraphics[width=8.6 cm]{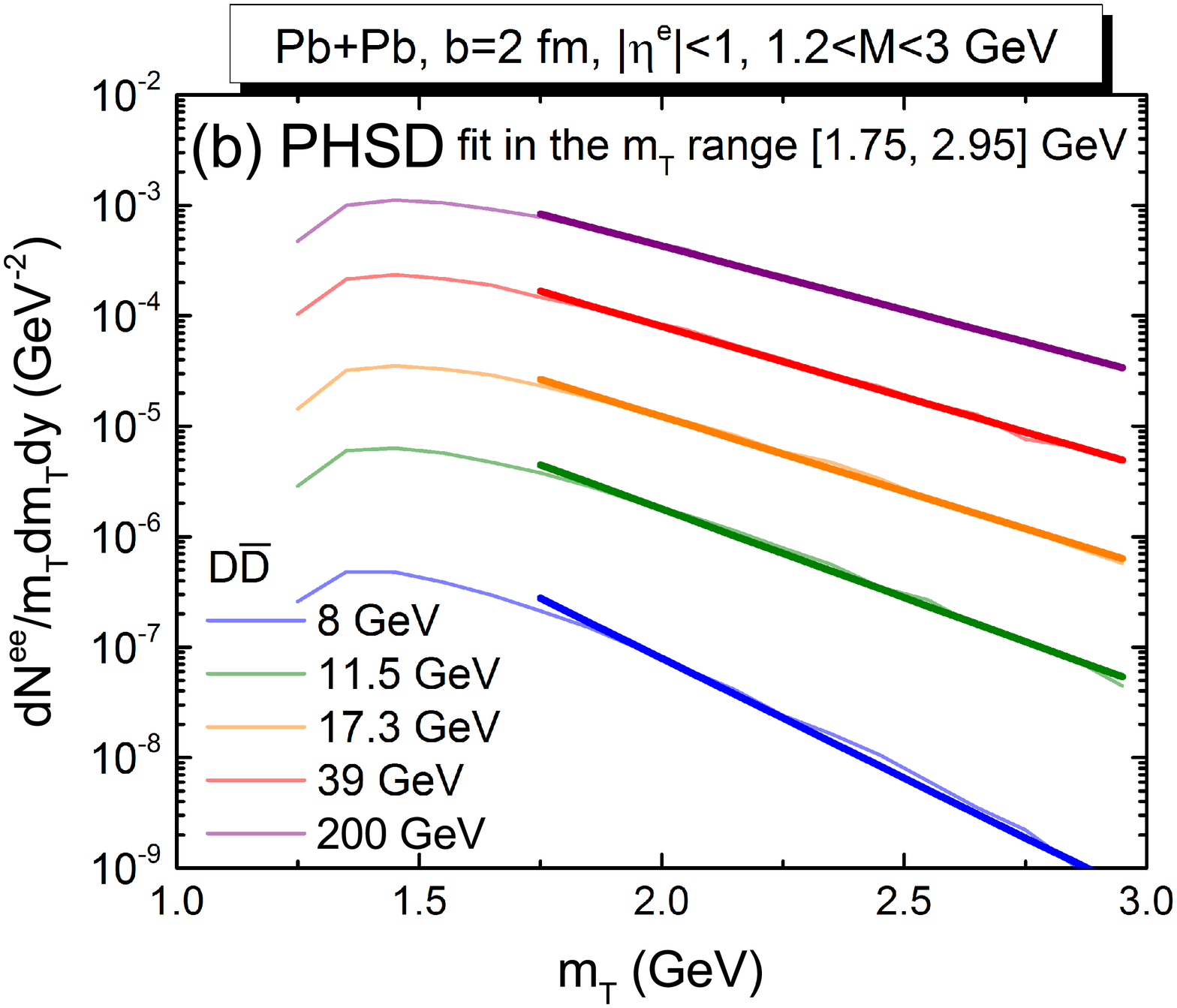}}
\centerline{
\includegraphics[width=8.6 cm]{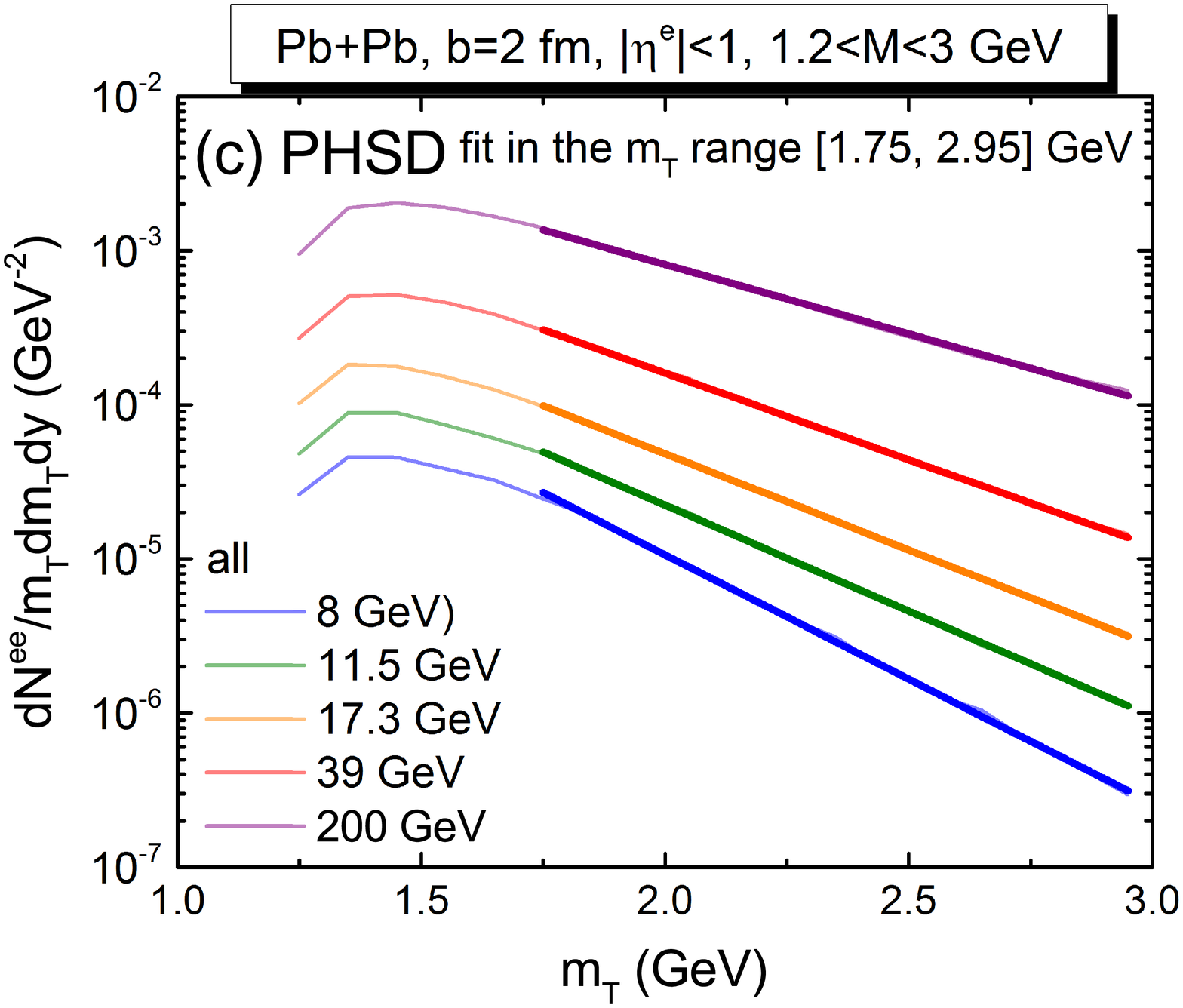}}
\caption{The transverse mass spectra of dileptons with the invariant mass between 1.2 and 3 GeV from the QGP (a), $D\bar{D}$ pairs (b), and all sources (c) in central Pb+Pb collisions at $\sqrt{s_{\rm NN}}$ = 8, 11.5, 17.3, 39 and 200 GeV from the PHSD. The fat solid lines show exponential fits to the PHSD results in the transverse mass range $[1.75, 2.95]$ GeV.} \label{spec}
\end{figure}

In this subsection we explore central Pb+Pb collisions at various energies with a focus on the transverse mass spectra of dileptons with intermediate-mass at midrapidity. To this aim we show in Fig.~\ref{spec} the Lorentz invariant transverse mass spectra for ($b$=2 fm) Pb+Pb collisions at $\sqrt{s_{\rm NN}}$ =  8, 11.5, 17.3, 39 and 200 GeV  for  the dielectrons with the invariant mass between 1.2 GeV and 3 GeV from the QGP (a), from $D$-mesons (b), and the dileptons from all channels (including especially $D, {\bar D}$ decay) (c). All spectra show an approximately exponential decay (fat solid lines) in the transverse mass $m_T$ for 1.75 GeV $< m_T <$ 2.95 GeV, which can be characterized by an inverse slope parameter $\beta$ which is different for dileptons from open charm and those from the QGP at all bombarding energies.

\begin{figure} [h]
\centerline{
\includegraphics[width=8.6 cm]{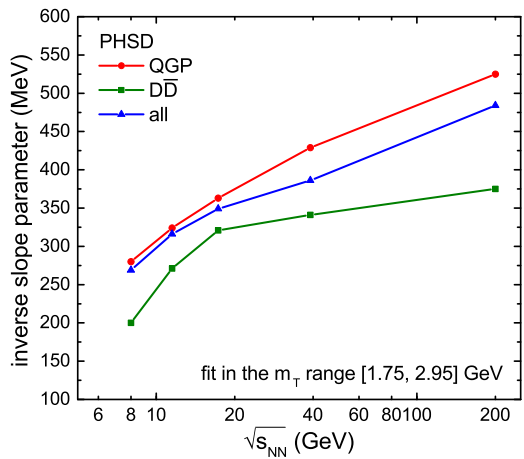}}
\caption{The inverse slope parameters of intermediate-mass dielectrons from the QGP (red line with dots), $D\bar{D}$ pairs (green line with squares), and all sources (blue line with triangles) in central Pb+Pb collisions at $\sqrt{s_{\rm NN}}$ = 8, 11.5, 17.3, 39 and 200 GeV from the PHSD.}
\label{temp}
\end{figure}

The excitation function in the inverse slope parameters $\beta$ is shown in Fig. \ref{temp} for the three cases of Fig. \ref{spec}, i.e. dileptons with the invariant mass between 1.2 and 3 GeV from the QGP (red line with dots), $D\bar{D}$ pairs (green line with squares), and all dilepton sources (blue line with triangles) in central Pb+Pb collisions at $\sqrt{s_{\rm NN}}$ = 8, 11.5, 17.3, 39 and 200 GeV. We find that the inverse slope parameter from the QGP contribution (red line with dots) is larger than the contribution from $D$-decays (green line with squares) at all energies and almost identical to the inverse slope for the total dilepton spectra (blue line with triangles) in the transverse mass range $[1.75, 2.95]$ GeV at SPS energies. Since the contribution from the $D$-decays increases with bombarding energy, a small wiggle in $\sqrt{s_{\rm NN}}$ can be found in the inverse slope for the total dilepton spectra (blue line with triangles) in the lower RHIC energy regime. This wiggle should be seen in experiment provided that high statistics data become available for intermediate mass dileptons.

\section{PHSD versus experimental data and predictions for the top LHC energy}
\subsection{Au+Au and Pb+Pb collisions from 19.6 GeV to 2.76 TeV}
In this section, we compare the invariant mass spectra of dielectrons from the PHSD to the experimental data in Au+Au collisions from $\sqrt{s_{\rm NN}}$ = 19.6 to 200 GeV from the STAR collaboration and those in Pb+Pb collisions from the ALICE collaboration at $\sqrt{s_{\rm NN}}$ = 2.76 TeV.
We note that the experimental data from the STAR collaboration and those from the ALICE collaboration have different centralites and different acceptance cuts.
The STAR data are obtained for minimum-bias Au+Au collisions and electrons and positrons with transverse momenta $p_T \geq$ 0.2 GeV and pseudo-rapidities $|\eta^e| < $ 1.0.
On other hand, the ALICE data are obtained for 0-10 \% central Pb+Pb collisions and the electrons and positrons with transverse momenta $p_T \geq$  0.4 GeV and  pseudo-rapidities $|\eta^e| < $ 0.8.
The sensitivity of the invariant mass spectra of dielectrons to the cross section for charm production and cuts in $p_T$ and pseudo-rapidity $\eta^e$ is discussed in more detail in Appendix B.

\begin{figure*}[h]
\centerline{
\includegraphics[width=8.6 cm]{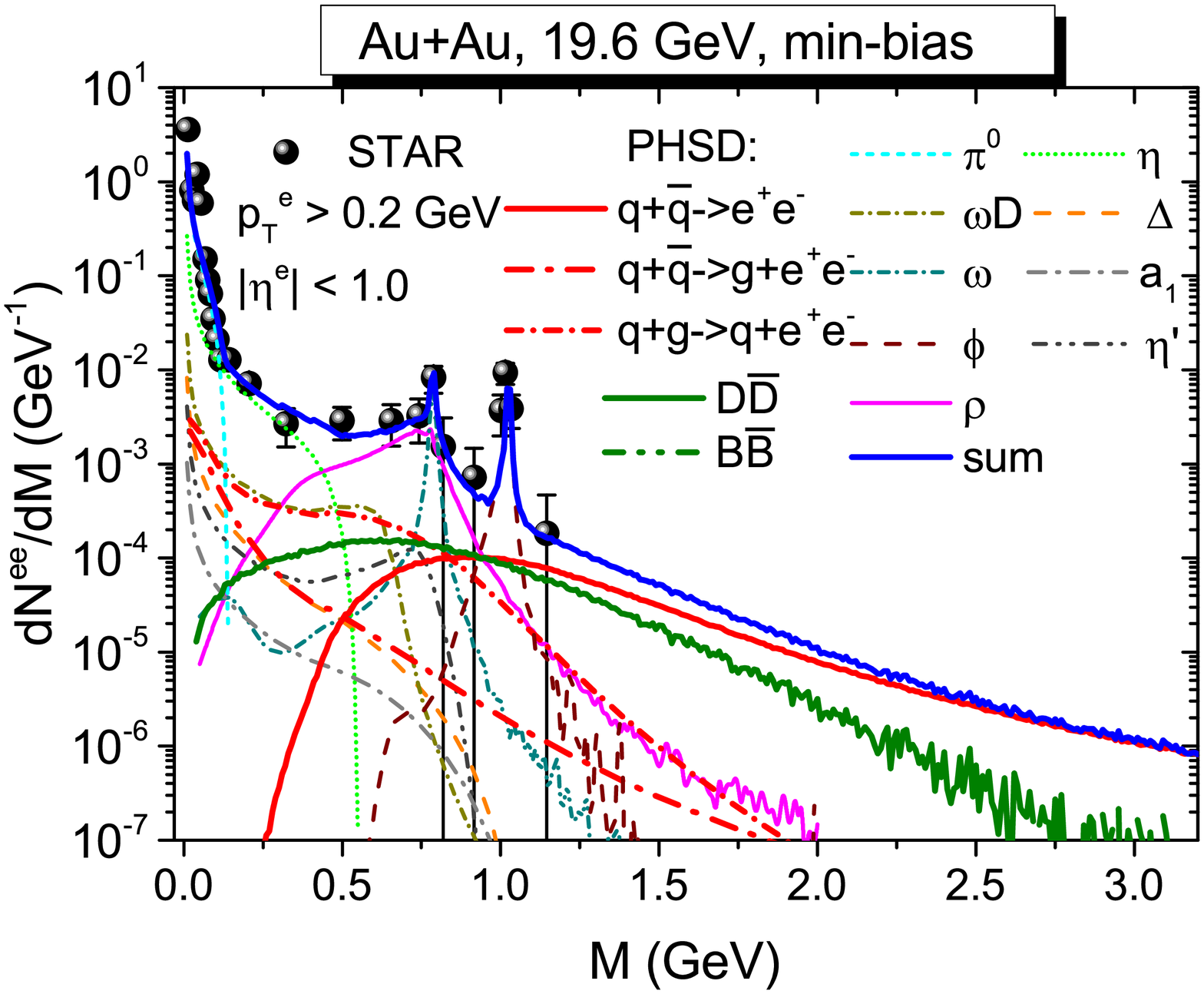}
\includegraphics[width=8.6 cm]{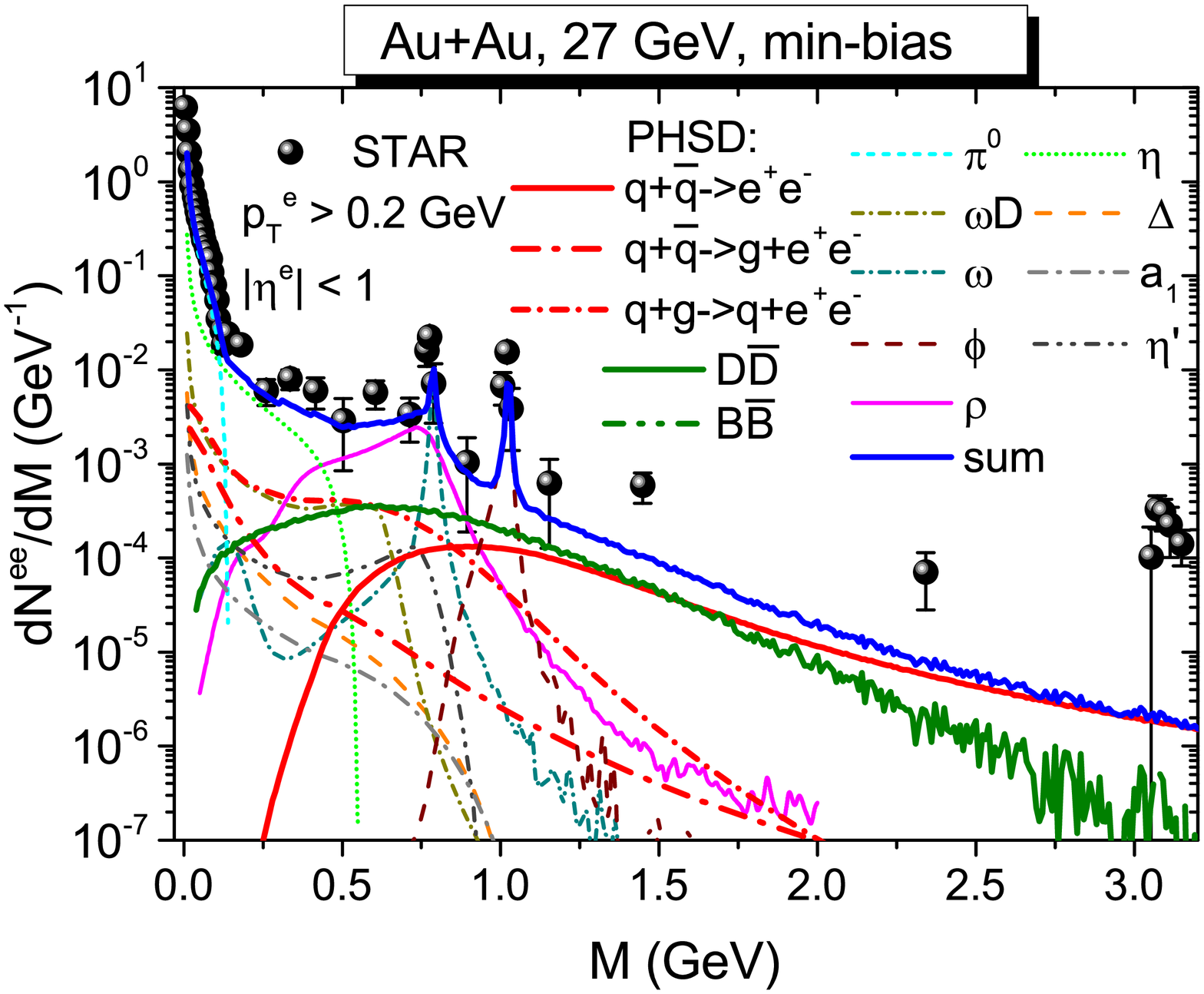}}
\centerline{
\includegraphics[width=8.6 cm]{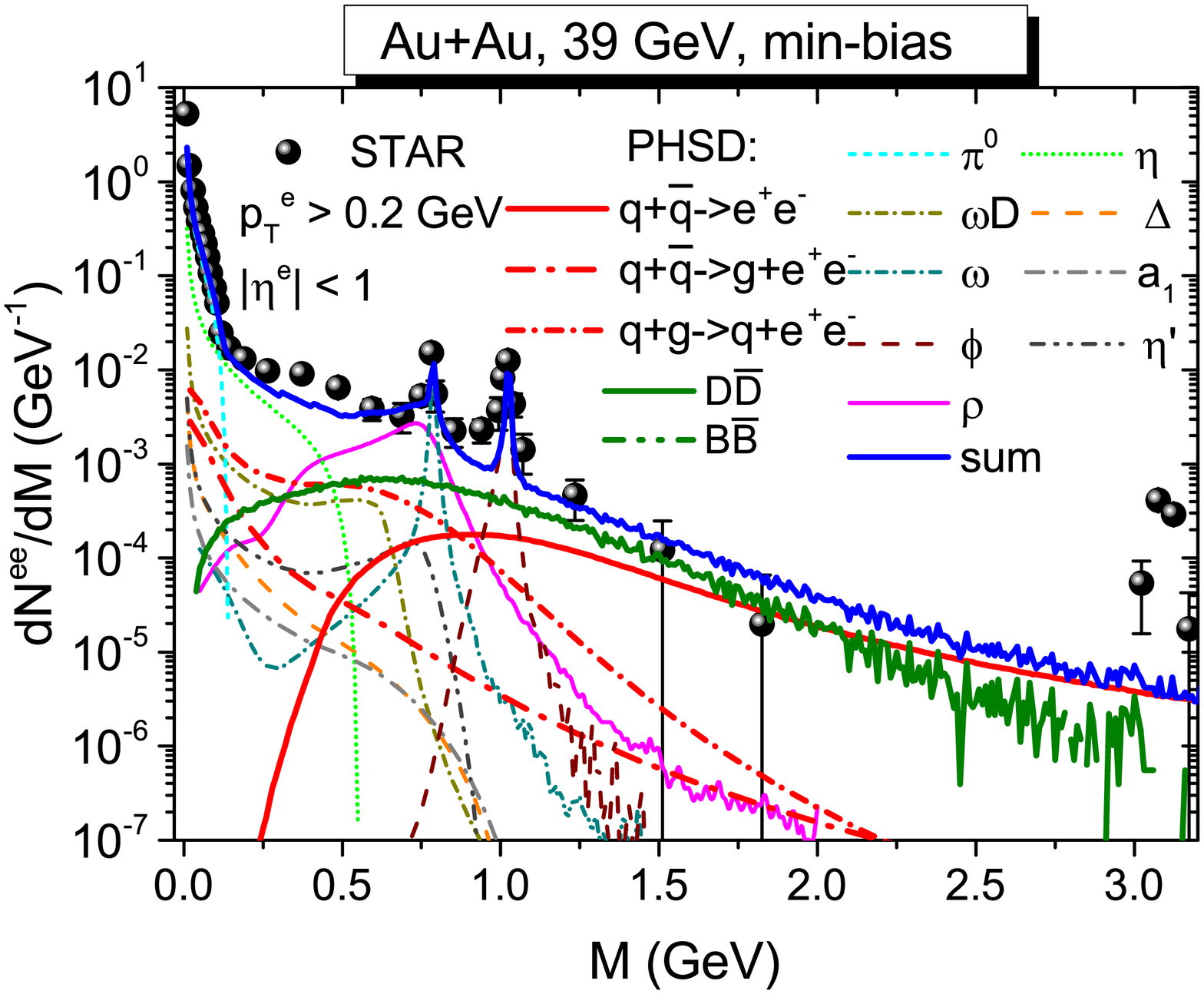}
\includegraphics[width=8.6 cm]{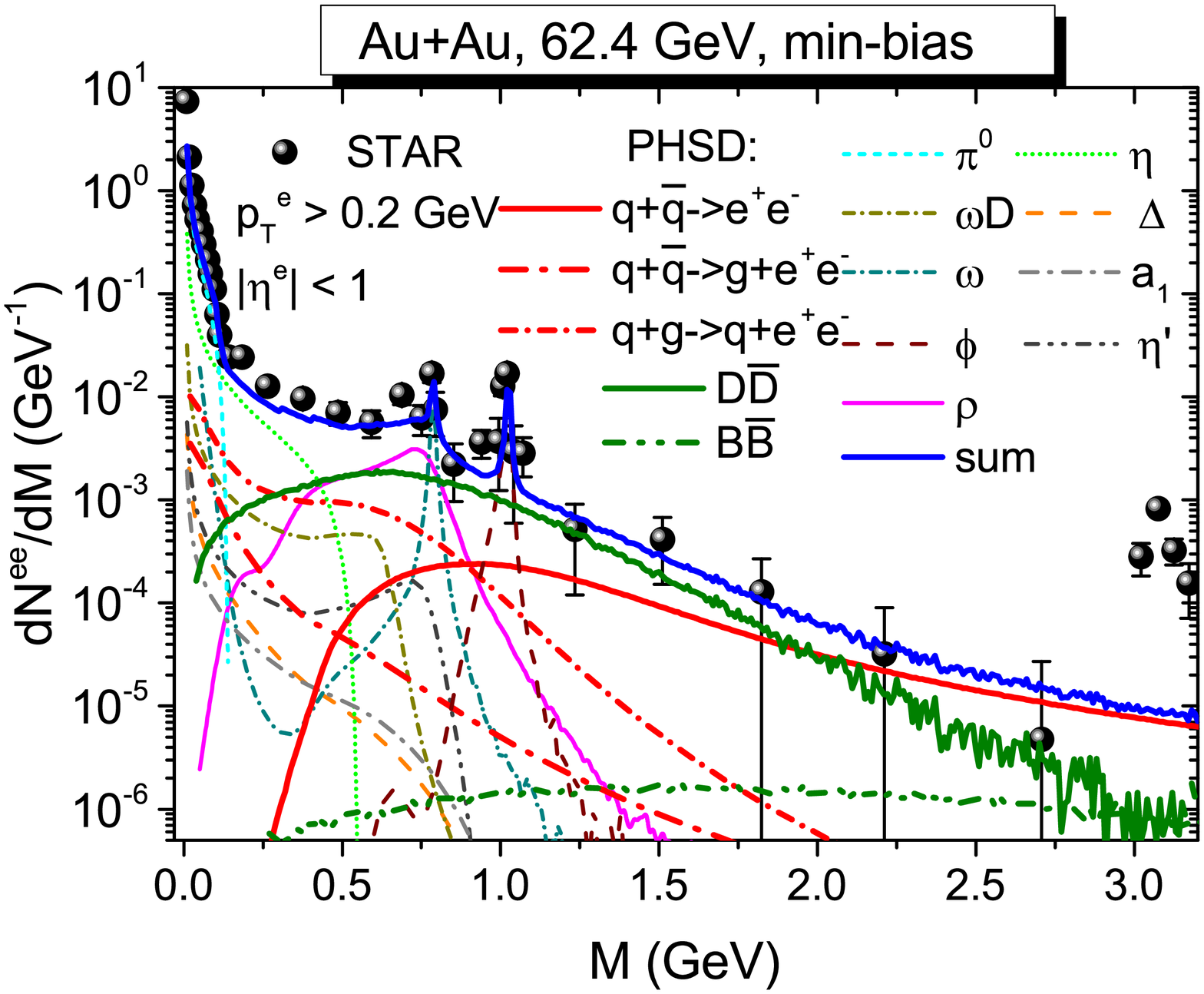}}
\centerline{
\includegraphics[width=8.6 cm]{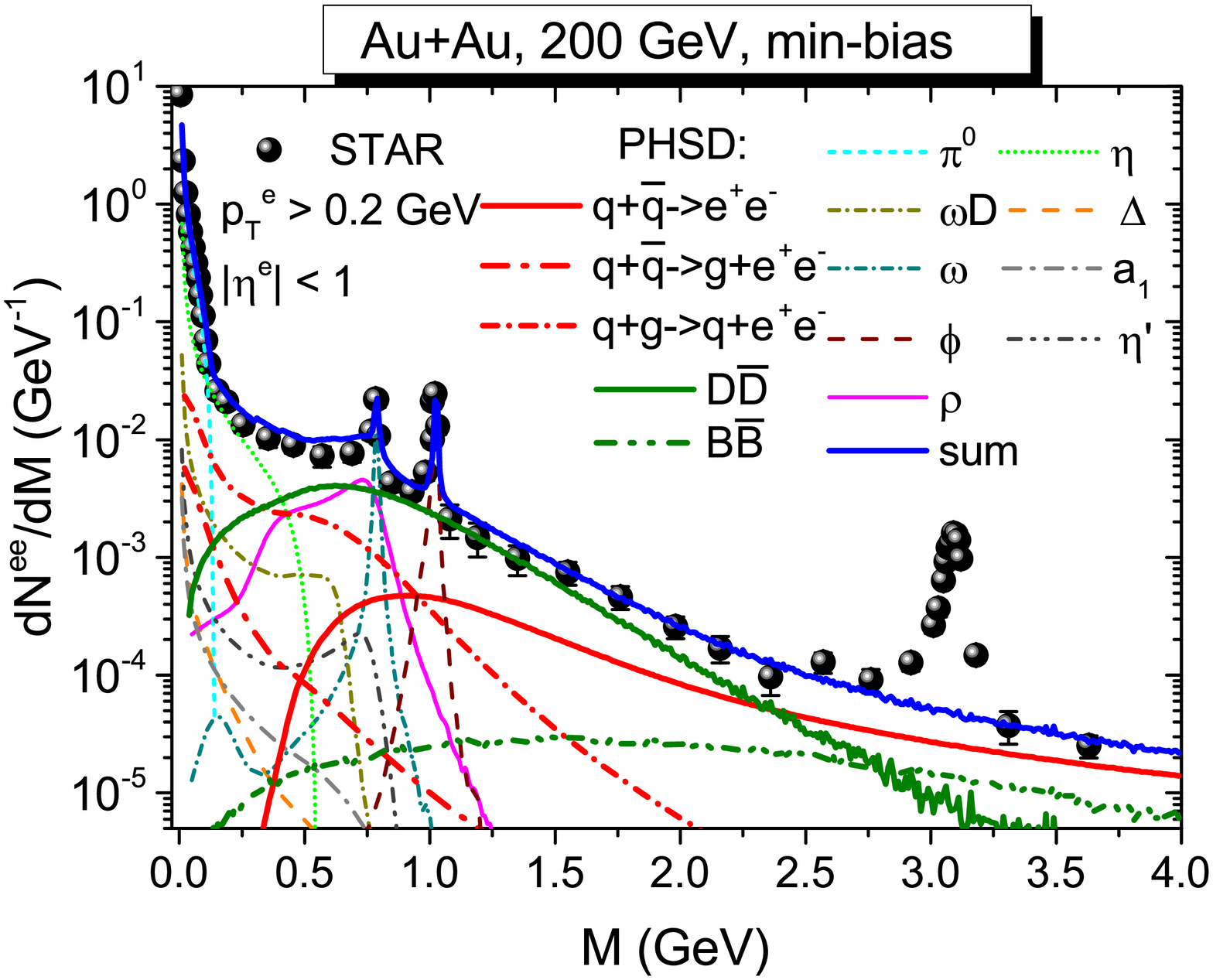}
\includegraphics[width=8.6 cm]{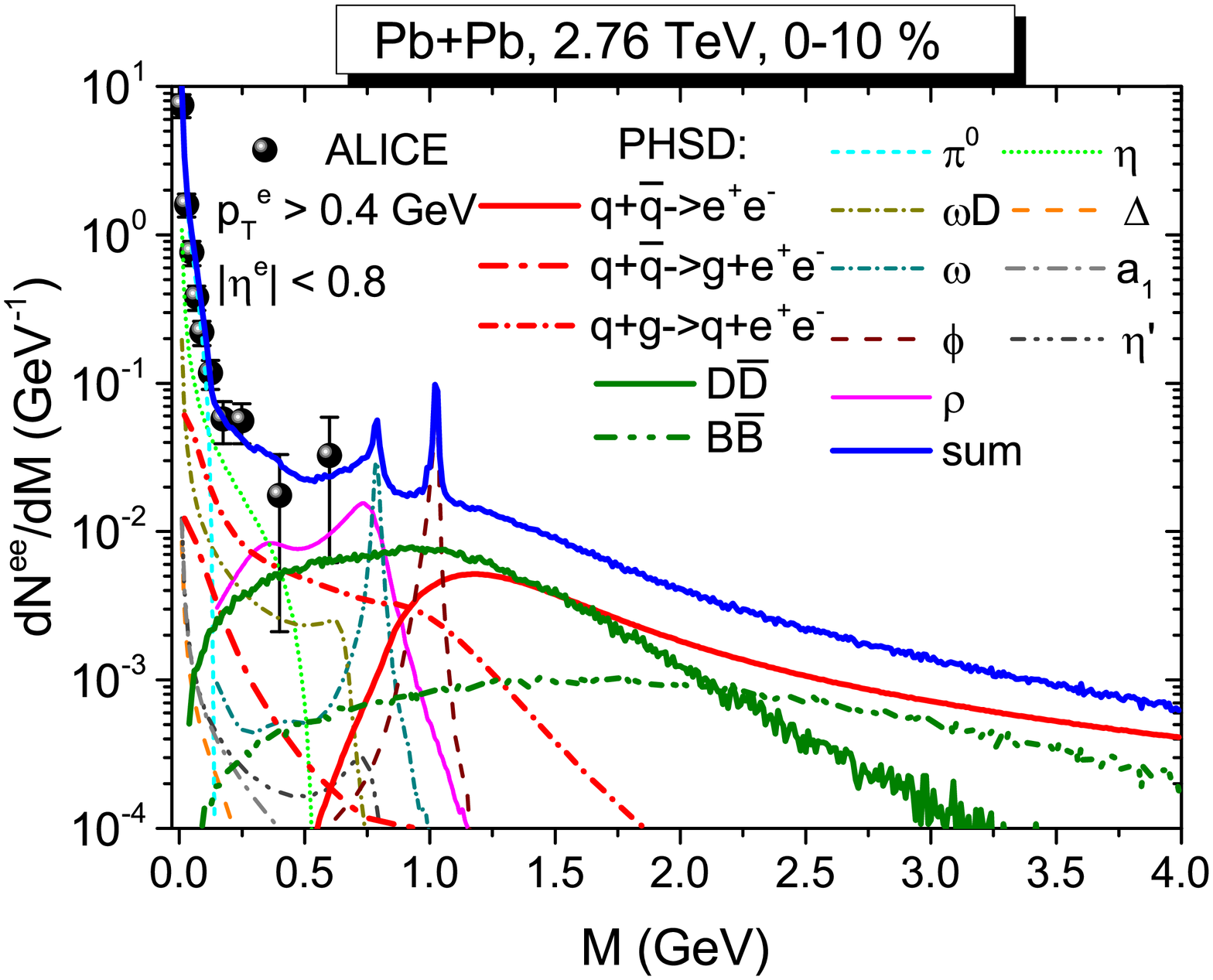}}
\caption{The invariant mass spectra of dielectrons from the PHSD in comparison to the STAR data in Au+Au collisions from $\sqrt{s_{\rm NN}}$ = 19.6 to 200 GeV~\cite{Huck:2014mfa,Adamczyk:2015lme} and to the ALICE data in Pb+Pb collisions at $\sqrt{s_{\rm NN}}$ = 2.76 TeV~\cite{Gunji:2017kot}. The total yield is displayed in terms of the blue lines while the different contributions are specified in the legends. Note that the contribution from $J/\Psi$ and $\Psi^\prime$ decays are not included in the PHSD calculations.}
\label{fig8}
\end{figure*}

The first five panels of Fig.~\ref{fig8} show the invariant mass spectra of dielectrons from the Beam-Energy-Scan (BES) at $\sqrt{s_{\rm NN}}$ = 19.6, 27, 39, and 62.4 GeV and from the top RHIC energy.
As discussed in the previous subsection, the contribution from hadrons is dominant in the low-mass region and signals  a broadening of the $\rho$ meson spectral function in dense nuclear matter (cf. Ref. \cite{PHSDreview}.
On the other hand, the intermediate-mass range originates predominantly by  dielectrons from partonic interactions and those from heavy flavor decays.
Similar to the Pb+Pb collisions in Fig.~\ref{fig6}, the contribution from heavy flavor becomes more and more important with increasing collision energy.
The contribution from heavy flavors and  from partonic interactions cross around invariant masses $M\approx$ 1 GeV in Au+Au collisions at $\sqrt{s_{\rm NN}}$ = 19.6 GeV.
However, the crossing point shifts to 1.6 GeV at $\sqrt{s_{\rm NN}}$ = 27 GeV and  to $\sim$2.0 GeV at $\sqrt{s_{\rm NN}}$ = 39 and 62.4 GeV.
At the top RHIC energy  they cross  at $\sim$2.4 GeV.

The last panel of Fig.~\ref{fig8} is the invariant mass spectrum of dielectrons in central Pb+Pb collisions at $\sqrt{s_{\rm NN}}$ = 2.76 TeV.
As in Au+Au collisions at the RHIC energies, the low-mass range is dominated by the dielecrons from hadronic channels and the intermediate-mass region by those from partonic interactions and heavy flavor decays.
However, the crossing point of the contribution from partonic interactions and that from heavy flavor is lower than at the top RHIC energy, which is due to a couple of effects:
i) the cross section for charm production  no longer increases rapidly at the LHC energies as shown in Fig.~\ref{fig1} (a). It is also seen in Fig.~\ref{fig1} (b), which shows the number of produced charm pairs as a function of collision energy. As a result, the growth in the number of produced charm pairs is not faster than the growth of dielectrons from partonic interactions.
Additionally the shadowing effect, which is the modification of the parton distribution function in nuclei~\cite{Eskola:2009uj}, considerably suppresses charm production at the LHC energies~\cite{Song:2015ykw} (see below).
ii) Another  reason is the stronger suppression of the charm four-momentum by partonic scattering at the LHC energies. As already discussed in the context of Fig.~\ref{fig4}, the strong interaction of heavy flavor with the medium reduces the invariant mass of dielectrons. Since the interaction is stronger at the LHC energies, we can expect a larger suppression of the dielectron spectrum at larger invariant masses.
iii) Furthermore,  at the LHC energies the contribution from semileptonic $B\bar{B}$ decays becomes important.
Comparing the lower two panels of Fig.~\ref{fig8}, the contribution from $B\bar{B}$ decays is found to be  larger than that from $D\bar{D}$ decays above $M \approx$ 2.2 GeV in Pb+Pb collisions at $\sqrt{s_{\rm NN}}$ = 2.76 TeV, while the contribution from $B\bar{B}$ decays is larger only above $M \approx$ 2.8 GeV in Au+Au collisions $\sqrt{s_{\rm NN}}$ = 200 GeV. Since the contribution from $B\bar{B}$ decays amounts to about 50\% of the contribution from partonic interactions at the LHC energies, it will distort the information on partonic matter in the intermediate-mass range of the dielectron spectrum.

Besides the interesting points mentioned above, we close this subsection with the comment that the dilepton invariant mass spectra from the PHSD describe reasonably well the available experimental data for collision energies from 19.6 GeV to 2.76 TeV although the experimental data at $\sqrt{s_{\rm NN}}$ = 2.76 TeV are available only for  invariant masses $M \leq$ 1 GeV.

\subsection{Predictions for central Pb+Pb {collisions} at $\sqrt{s_{\rm NN}}$= 5.02 TeV}

Based on the successful description of experimental data from the Beam Energy Scan for $\sqrt{s_{\rm NN}}$= 19.6 GeV to the LHC energy at $\sqrt{s_{\rm NN}}$= 2.76 TeV, we here provide predictions for dielectron production in central Pb+Pb collisions at  $\sqrt{s_{\rm NN}}$= 5.02 TeV.
As mentioned above, a proper description of heavy flavor production and interactions in heavy-ion collisions is necessary to allow for reliable predictions.

\begin{figure} [h]
\centerline{
\includegraphics[width=8.6 cm]{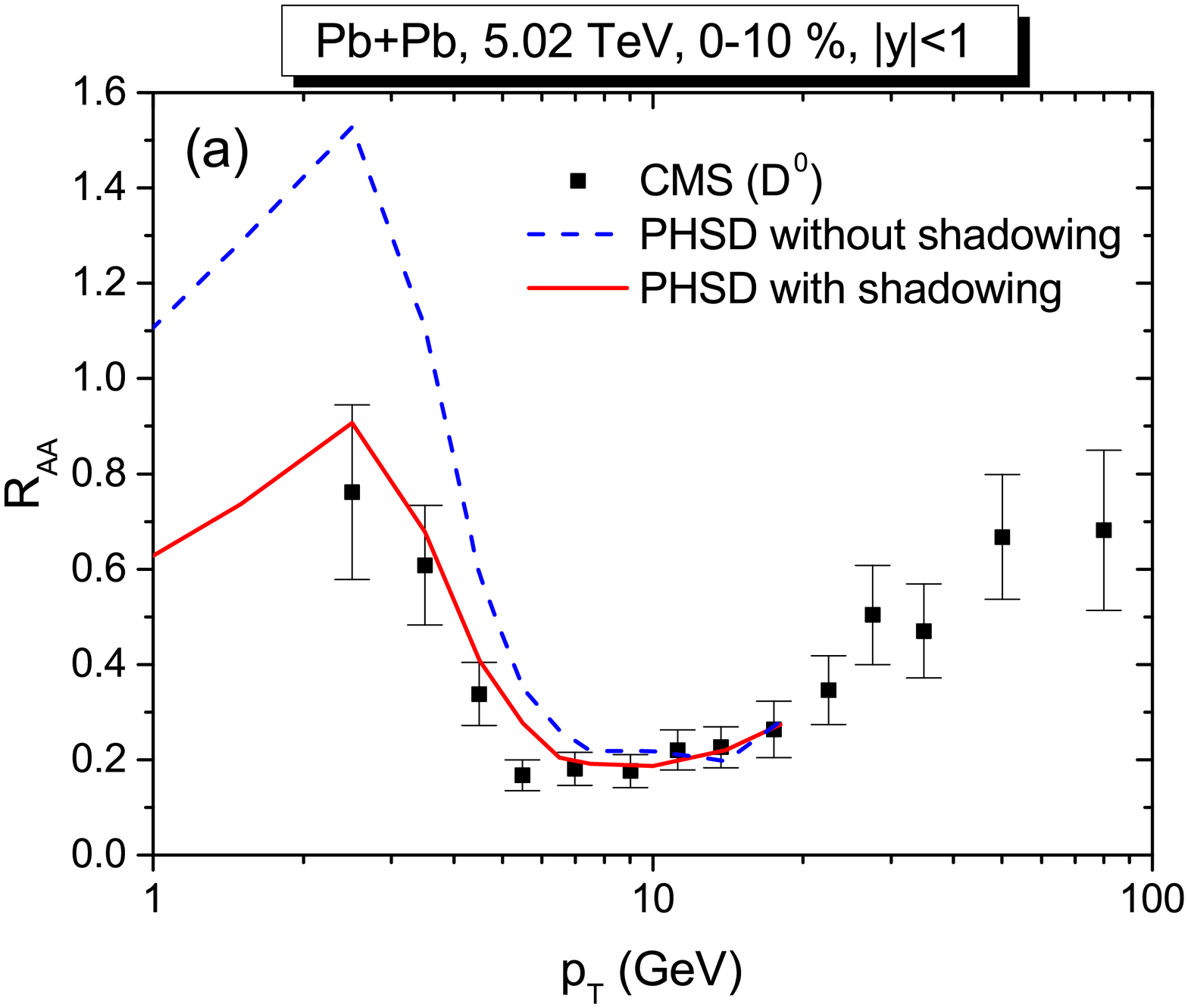}}
\centerline{
\includegraphics[width=8.6 cm]{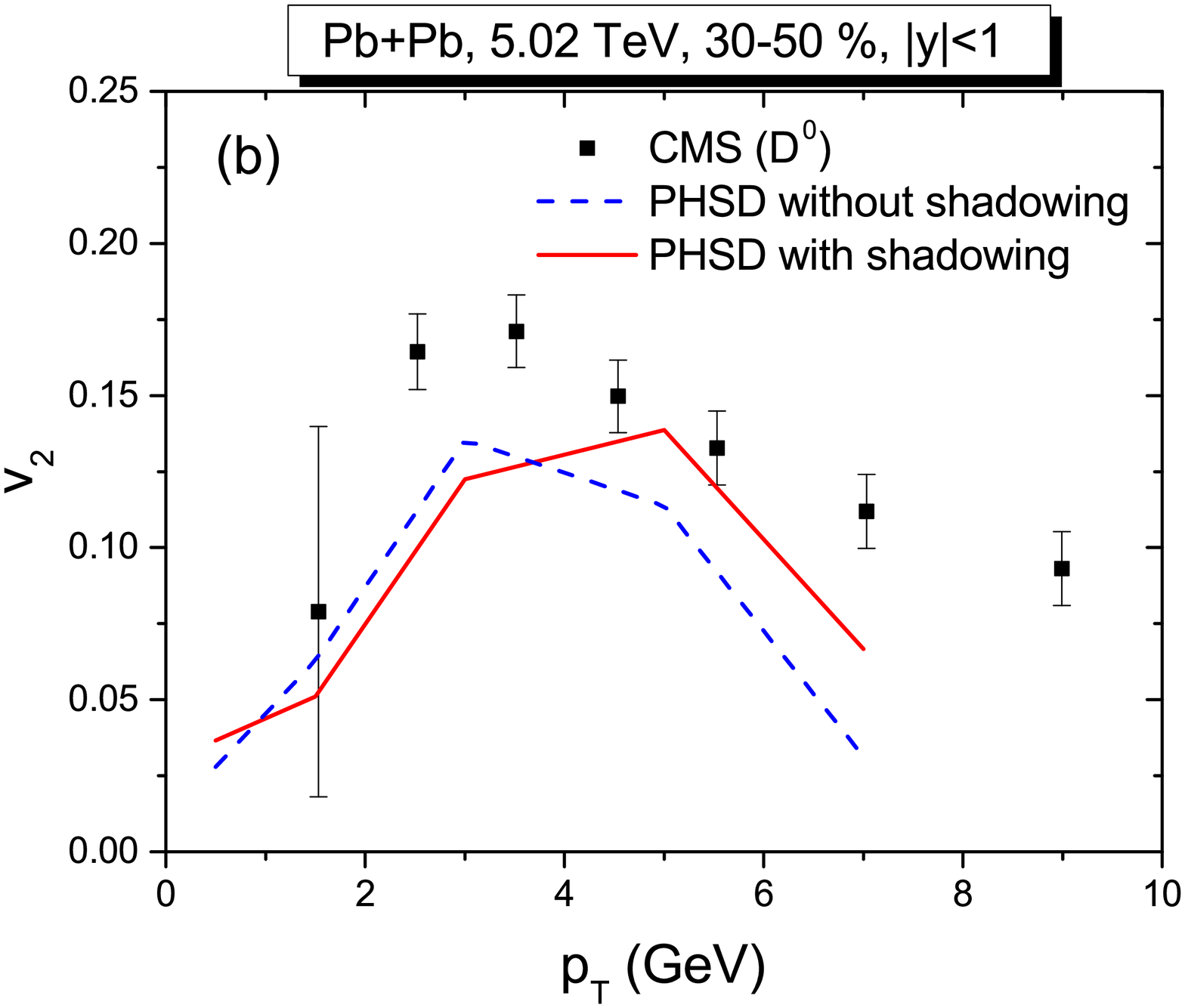}}
\caption{The $R_{AA}(p_T)$ (a) and  the elliptic flow $v_2(p_T)$ (b) for $D$ meson from 0-10 \% central Pb+Pb
collisions at $\sqrt{s_{\rm NN}}$= 5.02 TeV (from PHSD) as a function of the
transverse momentum with (solid line) and without shadowing effects
(dashed line). Experimental data are from the CMS collaboration~\cite{Sirunyan:2017xss,Sirunyan:2017plt}} \label{fig9}
\end{figure}

Fig.~\ref{fig9} shows the $R_{AA}$ (a) and the elliptic flow $v_2$ (b) of $D$ mesons as functions of transverse momentum in 0-10 \% central Pb+Pb collisions at $\sqrt{s_{\rm NN}}$= 5.02 TeV.
In both panels the dashed lines are the results without the shadowing effect and the solid lines with EPS09 shadowing \cite{Eskola:2009uj} included.
The upper panel shows that  shadowing reduces the $R_{AA}$ considerably at low transverse momentum, which can be explained as follws:
If the collision energy is very large, charm quark pairs with small transverse momentum are dominantly produced by partons with a small energy-momentum fraction $x$ of the nucleon.
On the other hand, the parton distribution function of a nucleon in a heavy nucleus is considerably suppressed at small $x$ in such  high-energy collisions~\cite{Eskola:2009uj}, which leads to a  suppression of charm production at low transverse momentum.
Fig.~\ref{fig9} (a) clearly shows that the shadowing effect is necessary to explain the experimental data from the ALICE collaboration.
We note that the PHSD results are presently available only up to $p_T=$ 20 GeV/c due to the limited statistics and huge CPU time required. In case of the open charm elliptic flow $v_2(p_T)$ the statistics do no allow for robust results for $p_T >$ 6 GeV/c. On the other hand, the shadowing effect is seen to have no substantial effect on the elliptic flow of $D$ mesons up to $p_T \approx$ 6 GeV/c since  shadowing  changes the production of charm from initial hard collisions but  does not change the interactions of produced charm in the partonic medium.
Fig.~\ref{fig9} demonstrates that both the $R_{AA}$ and the elliptic flow $v_2$ of $D$ mesons are approximately described at $\sqrt{s_{\rm NN}}$= 5.02 TeV by the PHSD.
Although the $v_2$ of $D$ mesons is slightly underestimated, this will have practically no effect on the dielectron spectrum.

\begin{figure} [h]
\centerline{
\includegraphics[width=8.6 cm]{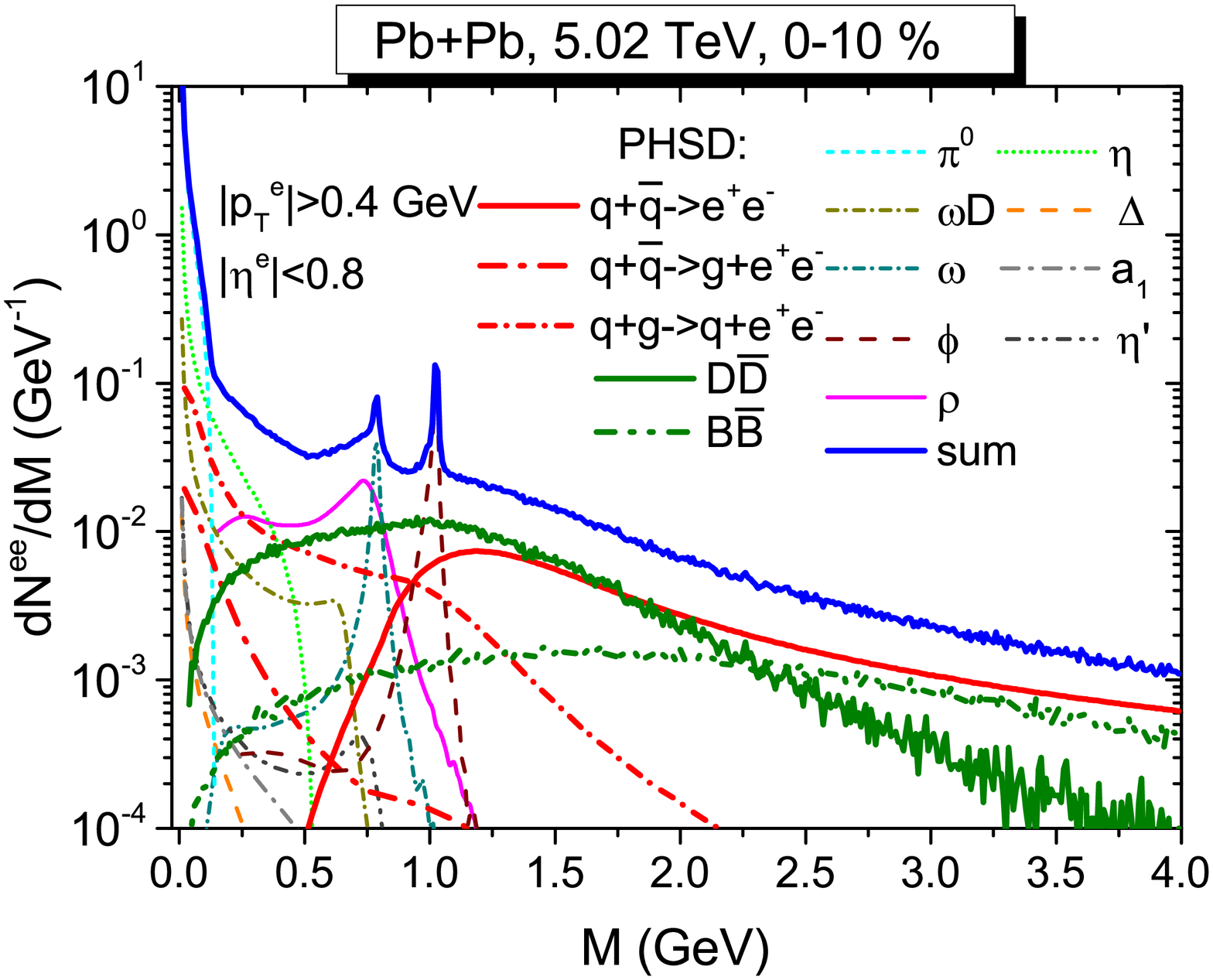}}
\caption{The invariant mass spectra of dielectrons for 0-10 \%
central Pb+Pb collisions at $\sqrt{s_{\rm NN}}$ = 5.02 TeV from the PHSD for $|p_T^e| > $ 0.4 GeV and $|\eta^e| <$ 0.8. }
\label{fig10}
\end{figure}

Fig.~\ref{fig10} shows the prediction from PHSD for the inva\-riant mass spectra of dielectrons in 0-10 \% central Pb+Pb collisions at $\sqrt{s_{\rm NN}}$ = 5.02 TeV
within the  acceptance  $(p_T>0.4 {\rm GeV},~ |\eta^e|<0.8)$ as in Fig.~\ref{fig8} (f).
Comparing with the results at 2.76 TeV we find no dramatic change in the shape of the spectrum except for an overall enhancement of the dielectron yield.
The yields of dielectrons from hadronic channels, from partonic interactions, and from heavy flavor decays are, respectively, enhanced by
55 \%, 54 \%, and 36 \% at $\sqrt{s_{\rm NN}}$ = 5.02 TeV.
We note that the dielectron yield from hadronic channels  and that from partonic interactions increase by a similar amount suggesting that both dielectron yields are produced from bulk matter whereas the
dielectron yield from heavy flavor decays is less enhanced due to a lower increase in the charm production cross section.

\section{Summary}\label{summary}
We have studied correlated electron ($e^+ e^-$) production through
the semileptonic decay of charm hadrons in relativistic heavy-ion
collisions from $\sqrt{s_{\rm NN}}=$ 8 GeV to 5 TeV within the PHSD
transport approach in extension of our work on  $D-$meson production
in relativistic heavy-ion collisions at RHIC and LHC
energies~\cite{Song:2015sfa,Song:2015ykw,Song2016} and low mass dilepton
production from SIS to RHIC energies \cite{PHSDreview}.

In the PHSD the  charm  partons - produced by the initial hard
nucleon-nucleon scattering - interact with the massive quarks and
gluons in the QGP by using the scattering cross sections calculated
in the Dynamical Quasi-Particle Model (DQPM) which reproduces
heavy-quark diffusion coefficients from lattice QCD calculations at
temperatures above the deconfinement transition. When approaching
the critical energy density for the phase transition from above, the
charm  (anti)quarks are hadronized into $D-$mesons through the
coalescence with light (anti)quarks. Those heavy quarks, which fail
in coalescence until the local energy density is below 0.4 $\rm
GeV/fm^3$, hadronize by fragmentation as in p+p collisions. The
hadronized $D-$mesons then interact with light hadrons in the
hadronic phase with cross sections that have been calculated in an
effective lagrangian approach with heavy-quark spin symmetry.
Finally, after freeze-out of the $D-$mesons they produce single
electrons through semileptonic decays with the branching ratios
given by the  PYTHIA event generator.

The dilepton production from hadronic and
partonic channels in central Pb+Pb (or Au+Au) collisions has been calculated including
also the contribution from the semileptonic decays of heavy flavors in
PHSD for the first time on a fully microscopic level.
We recall that also the cross sections for dilepton production have been calculated by employing the same propagators and couplings as incorporated in the partonic dynamics in PHSD (cf. Appendix A).
We find that even in central Pb+Pb
collisions at $\sqrt{s_{\rm NN}}=$ {8} to 20 GeV the contribution from
$D,{\bar D}$ mesons to the intermediate mass dilepton spectra is
subleading and one should have a rather clear signal from the QGP
radiation whereas at the top RHIC energy this contribution
overshines the intermediate mass dileptons from the QGP.
{It is interesting to note that the dielectrons from $D,{\bar D}$ mesons do not increase any more relative to  partonic interactions at the LHC energies for a couple of reasons:
i) the cross section for charm production does not grow as fast as at low energies;
ii) the shadowing effects, which suppress charm production at low transverse momentum, are stronger at LHC than at RHIC energies (cf. Fig. 11);
iii) the  charm quark pair looses more four-momentum in the partonic medium produced at the LHC, which suppresses the invariant mass of the dielectrons from the semileptonic decays.
Furthermore, the contribution from $B,{\bar B}$ meson decays becomes more important and
superseeds the contribution from $D,{\bar D}$ meson decays above $M=$ 2.2$\sim$2.3 GeV at the LHC energies and amounts to about half the contribution from partonic interactions.
All these effects strongly distort the information about partonic matter from intermediate-mass dielectrons at the LHC energies.} The dilepton spectra at lower masses ($0.2 ~{\rm GeV} \leq M \leq
0.7 ~{\rm GeV})$ at SPS, FAIR/NICA and BES RHIC energies show some sensitivity to the medium
modification of the $\rho$ meson where the data favor an in-medium
broadening as pointed out in the earlier studies on dilepton production reviewed in Refs. \cite{ChSym,PHSDreview}.

Additionally, we have explored the transverse mass spectra of dileptons in the invariant mass range from 1.2 GeV to 3 GeV in central Pb+Pb collisions for $\sqrt{s_{\rm NN}}$ = 8 to 200 GeV and find  approximately exponential spectra for transverse masses in the energy range $[1.75, 2.95]$ GeV (cf. Fig. 8). Since the inverse slope parameters differ for the contributions from the QGP and are higher than that from $D$-decays we expect a wiggle in the excitation function of the inverse slope parameter for these intermediate mass dileptions (cf. Fig. 9) which should be seen experimentally in high statistics data.

In general the PHSD calculations compare well with the available dilepton data from the BES program at RHIC as well as the LHC energy of $\sqrt{s_{\rm NN}}$ = 2.76 TeV where, unfortunately, only low mass dilepton data are available so far.
Explicit predictions for central Pb+Pb collisions at $\sqrt{s_{\rm NN}}$ = 2.76 TeV have been provided (cf. Fig. 12),
however, the partonic contribution in the intermediate mass range has a large background from $D, {\bar D}$  as
well as $B, {\bar B}$ correlated semi-leptonic decays. As noted above, this  background - in the intermediate mass range - is by far subleading at lower SPS and FAIR/NICA energies which provides promising perspectives for the future dilepton measurements at these facilities and allows for a fresh look at the electromagnetic radiation from the QGP
at finite baryon chemical potential.

\section*{Acknowledgements}
The authors acknowledge inspiring discussions with
J. Butterworth, F. Geurts and O. Linnyk.
This work was supported by the LOEWE center "HIC for FAIR", the HGS-HIRe for FAIR and
the COST Action THOR, CA15213.
Furthermore, PM and EB acknowledge support by DFG through the grant CRC-TR 211 'Strong-interaction matter under extreme conditions'.
The computational resources have been provided by the LOEWE-CSC.


\hfil\break
\appendix
\bigskip

\section{}
\subsection{Leading order contribution}

In this Appendix, we provide the details on the cross sections for the processes $q+\bar{q}\rightarrow \gamma^*(e^+e^-)$, $q+g\rightarrow q+\gamma^*(e^+e^-)$, and $q+\bar{q}\rightarrow g+\gamma^*(e^+e^-)$, considering the off-shellness of the interacting partons in line with the DQPM.

\begin{figure}[h]
\centering
{\includegraphics[width=5 cm]{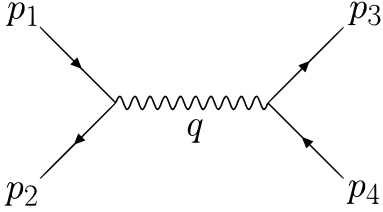}}
\caption{Feynman diagram for $q+\bar{q}\rightarrow l^++l^-$}
\label{feynman-lo}
\end{figure}

The invariant matrix element for the process $q+\bar{q}\rightarrow l^++l^-$ is given by
\begin{eqnarray}
M=\bar{u}(p_3)ie\gamma^\nu v(p_4)\frac{-ig_{\mu\nu}}{q^2}\bar{v}^a(p_2)ig\gamma^\mu \delta_{ab}u^b(p_1),
\end{eqnarray}
where $a$ and $b$ are the color indices of the incoming quark and antiquark. The matrix element squared then reads as
\begin{eqnarray}
|\overline{M}|^2=\frac{8(4\pi\alpha)^2}{N_c}\frac{1}{s^2}\{(p_1\cdot p_3)(p_2\cdot p_4) +(p_1\cdot p_4)(p_2\cdot p_3)\nonumber\\
+m_1m_2(p_3\cdot p_4)+m_l^2(p_1\cdot p_2)+2m_l^2m_1m_2\},~~~~~
\end{eqnarray}
with $N_c=3$ for the number of colors. Here, $m_1$, $m_2$ and $m_l$ are, respectively, quark and antiquark masses and lepton mass, and the following color algebra is used:
\begin{eqnarray}
|\overline{M}|^2 \sim \frac{1}{N_c^2}\bar{v}^a(p_2)\delta_{ab}u^b(p_1)\bar{u}^{b^\prime}(p_1)\delta_{a^\prime b^\prime}v^{a^\prime}(p_2) \nonumber\\
\sim \frac{1}{N_c^2}\delta_{a a^\prime}\delta_{ab}\delta_{a^\prime b^\prime}\delta_{b b^\prime}=\frac{1}{N_c^2}\delta_{aa}=\frac{1}{N_c}.
\end{eqnarray}
We note that $m_1$ is not necessarily equal to $m_2$ since the masses of the incoming quark and antiquark have spectral distributions which depend on the local temperature as defined by the DQPM.
The phase-space integration for the scattering cross section is straightforward (cf. Ref.~\cite{Agashe:2014kda}).

\subsection{Next-to-leading order contributions}

\begin{figure}[h]
\centering
{\includegraphics[width=0.45\textwidth]{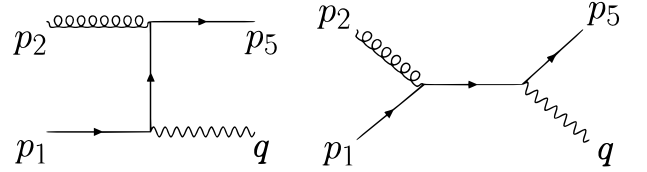}}
\caption{Feynman diagrams for the process $q+g\rightarrow q+\gamma^*$}
\label{feynman1}
\end{figure}

The invariant matrix element for the process $q+g\rightarrow q+\gamma^*(e^+e^-)$ in Fig.~\ref{feynman1} reads,
\begin{eqnarray}
M=\bar{u}_b(p_5)\bigg[T^a_{bc}\frac{ig\gamma^\alpha i(\not{p_5}-\not{p_2}+m_q)\delta_{cd}ie_q\gamma^\beta}{(p_2-p_5)^2-m_q^2+i2(p_{20}-p_{50})\Gamma_q} \nonumber\\
+\frac{\delta_{bc}ie_q\gamma^\beta i(\not{p_1}+\not{p_2}+m_q)ig\gamma^\alpha }{(p_1+p_2)^2-m_q^2+i2(p_{10}+p_{20})\Gamma_q}T^a_{cd}\bigg]u_d(p_1)\nonumber\\
\times \varepsilon_\alpha^{a*}(p_2)L_\beta~~~~~,
\label{invariantm1}
\end{eqnarray}
where $p_1$, $p_2$, and $p_5$ are the momenta of incoming quark, incoming gluon and outgoing quark with $p_{10}$, $p_{20}$, and $p_{50}$ denoting their zeroth components, $d$, $a$, and $b$ are their color indices, $\alpha$ and $\beta$ are spin indices of the incoming gluon and outgoing photon, $e_q$ is the electric charge of the quark, $m_q$ and $\Gamma_q$ are, respectively, the mass and the spectral width of the intermediate quark, and $L_\beta$ is defined as
\begin{eqnarray}
L_\beta=\bar{u}_b(p_3)ie\gamma_\beta v(p_4)/(iq^2),
\end{eqnarray}
with $p_3$ and $p_4$ denoting the momenta of electron and positron and $q=p_3+p_4$.
In the case of real photon production $L_\beta$ is replaced by $\varepsilon_\alpha(q)$.
The first term in Eq.~(\ref{invariantm1}) is the u-channel and the second term the s-channel.
We note that the imaginary part of the quark propagator in the u-channel is taken proportional to $p_{20}-p_{50}$ (downward) rather than to $p_{50}-p_{20}$ (upward), because the gluon mass is roughly twice the quark mass.

The invariant matrix element squared is written as follows:
\begin{widetext}
\begin{eqnarray}
|M|^2=8(4\pi)^2\alpha\alpha_s\bigg(\frac{e_q}{e}\bigg)^2\frac{N_c^2-1}{2}L_\beta L_{\beta^\prime}^*\bigg[~~~~~~~~~~~~~~~~~~~~~~~~~~~~~~~~~~~~~~~~~~~~~~~~~~~~~~~~~~~~~~~~~\nonumber\\
\bigg(\{4m_5m_q- 2p_5\cdot(p_5-p_2)\}\{p_1^\beta(p_5-p_2)^{\beta^\prime}+p_1^{\beta^\prime}(p_5-p_2)^\beta -p_1\cdot(p_5-p_2)g^{\beta \beta^{\prime}}\}\nonumber\\
+\{(p_5-p_2)^2-m_q^2\}\{p_5^\beta p_1^{\beta^\prime} +p_5^{\beta^\prime}p_1^\beta -p_5\cdot p_1 g^{\beta \beta^{\prime}}\}\nonumber\\
+2m_1 g^{\beta \beta^{\prime}}\{m_5(p_5-p_2)^2 +m_5m_q^2 -m_q p_5\cdot(p_5-p_2)\}\bigg)\frac{-1}{\{(p_2-p_5)^2-m_q^2\}^2 +4(p_{20}-p_{50})^2\Gamma_q^2}\nonumber\\
+\bigg(\{4m_1m_q- 2p_1\cdot(p_1+p_2)\}\{p_5^\beta(p_1+p_2)^{\beta^\prime}+p_5^{\beta^\prime}(p_1+p_2)^\beta -p_5\cdot(p_1+p_2)g^{\beta \beta^{\prime}}\}\nonumber\\
+\{(p_1+p_2)^2-m_q^2\}\{p_5^\beta p_1^{\beta^\prime} +p_5^{\beta^\prime}p_1^\beta -p_5\cdot p_1 g^{\beta \beta^{\prime}}\}\nonumber\\
+2m_5 g^{\beta \beta^{\prime}}\{m_1(p_1+p_2)^2 +m_1m_q^2 -m_q p_1\cdot(p_1+p_2)\}\bigg)\frac{-1}{\{(p_1+p_2)^2-m_q^2\}^2 +4(p_{10}+p_{20})^2\Gamma_q^2}\nonumber\\
-2\bigg(-(p_5\cdot p_1)\{(p_5-p_2)^\beta(p_1+p_2)^{\beta^\prime} +(p_5-p_2)\cdot(p_1+p_2)g^{\beta\beta^\prime}-(p_5-p_2)^{\beta^\prime}(p_1+p_2)^\beta\}\nonumber\\
+p_5^\beta\{p_1\cdot(p_5-p_2)(p_1+p_2)^{\beta^\prime} +(p_5-p_2)\cdot(p_1+p_2)p_1^{\beta^\prime} -(m_1^2+p_1\cdot p_2)(p_5-p_2)^{\beta^\prime}\}\nonumber\\
-(m_5^2-p_5\cdot p_2)\{p_1^\beta(p_1+p_2)^{\beta^\prime} +p_1^{\beta^\prime}(p_1+p_2)^\beta -(m_1^2 +p_1\cdot p_2)g^{\beta\beta^\prime}\}\nonumber\\
+\{p_5\cdot(p_1+p_2)-m_5m_q\}\{p_1^\beta(p_5-p_2)^{\beta^\prime} +p_1^{\beta^\prime}(p_5-p_2)^\beta -p_1\cdot (p_5-p_2)g^{\beta\beta^\prime}\}\nonumber\\
-p_5^{\beta^\prime}\{(p_5-p_2)\cdot(p_1+p_2)p_1^\beta +(m_1^2+p_1\cdot p_2)(p_5-p_2)^\beta -p_1\cdot(p_5-p_2)(p_1+p_2)^\beta\}\nonumber\\
-m_qm_1\{p_5^\beta(p_1+p_2)^{\beta^\prime} +p_5^{\beta^\prime}(p_1+p_2)^\beta -p_5\cdot(p_1+p_2)g^{\beta\beta^\prime}\}\nonumber\\
+2m_5 m_q p_1^\beta(p_1+p_2)^{\beta^\prime} + 2m_5 m_1 (p_5-p_2)^\beta (p_1+p_2)^{\beta^\prime} +2m_q^2 p_1^\beta p_5^{\beta^\prime}\nonumber\\
+2m_1 m_q (p_5-p_2)^\beta p_5^{\beta^\prime} - m_5 m_q^2 m_1 g^{\beta\beta^\prime}\bigg)\nonumber\\
\times\frac{\{(p_2-p_5)^2-m_q^2\}\{(p_1+p_2)^2-m_q^2\}+4(p_{10}+p_{20})(p_{20}-p_{50})\Gamma_q^2}
{\bigg((p_2-p_5)^2-m_q^2\}^2 +4(p_{20}-p_{50})^2\Gamma_q^2\bigg)\bigg(\{(p_1+p_2)^2-m_q^2\}^2 +4(p_{10}+p_{20})^2\Gamma_q^2\bigg)}~\bigg],
\label{invariantm0}
\end{eqnarray}
\end{widetext}
where it is assumed that $\varepsilon_\alpha^{a*}(p_2)\varepsilon_{\alpha^\prime}^{a^\prime}(p_2)=-g_{\alpha\alpha^\prime}\delta_{aa^\prime}$ according to the Lorentz gauge, and
\begin{eqnarray}
L^\beta L^{\beta^\prime*}=16\pi\alpha\frac{p_3^\beta p_4^{\beta^\prime} +p_4^\beta p_3^{\beta^{\prime}} -g^{\beta{\beta^\prime}}(q^2/2)}{q^4}.
\label{dileptons}
\end{eqnarray}
We note that the nonvanishing width of the quark spectral function removes divergences which appear in some kinetic regions.
The first three lines in the bracket of Eq.~(\ref{invariantm0}) is the squared u-channel and the next three lines the squared s-channel and the rest the mixed term of u-channel and t-channel.
We note that the squared u-channel and the squared s-channel are equivalent to each other, if $p_1$ and $-p_5$, and $m_1$ and $m_5$ are exchanged.

\begin{figure}[h]
\centering
{\includegraphics[width=0.45\textwidth]{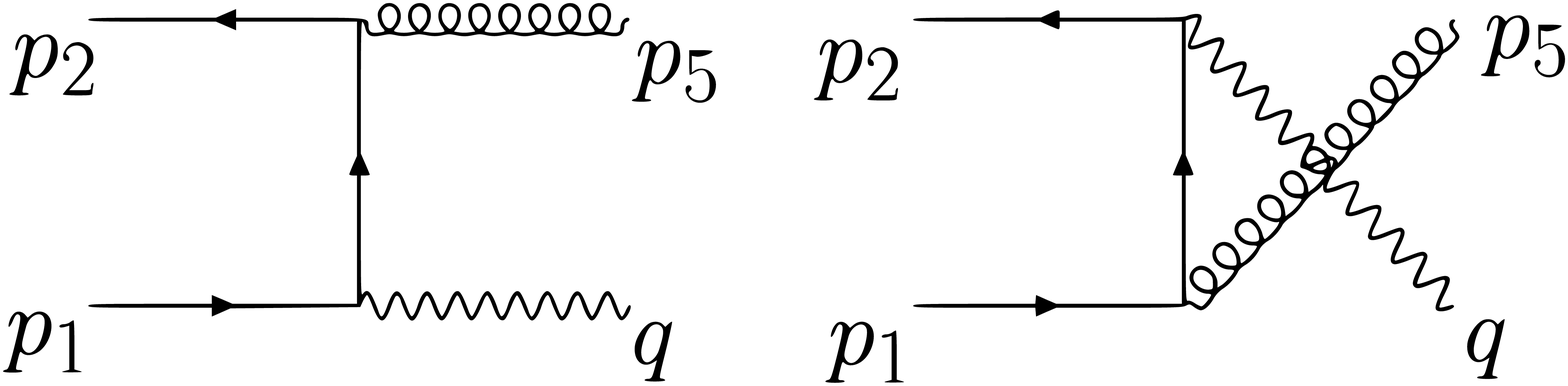}}
\caption{Feynman diagrams for $q+\bar{q}\rightarrow g+\gamma^*$}
\label{feynman2}
\end{figure}

The invariant matrix element for the process $q+\bar{q}\rightarrow g+\gamma^*$ is calculated from  Fig.~\ref{feynman2} as
\begin{eqnarray}
M=\bar{v}_d(p_2)\bigg[\frac{\delta_{db}ie_q\gamma^\beta i(\not{p_1}-\not{p_5}+m_q)ig\gamma^\alpha}{(p_1-p_5)^2-m_q^2+i2(p_{10}-p_{50})\Gamma_q} T^a_{bc}\nonumber\\
+T^a_{db}\frac{ig\gamma^\alpha i(-\not{p_2}+\not{p_5}+m_q)\delta_{bc}ie_q\gamma^\beta}{(-p_2+p_5)^2-m_q^2+i2(-p_{20}+p_{50})\Gamma_q}\bigg]u_c(p_1)\nonumber\\
\times\varepsilon_\alpha^{a}(p_5)L_\beta~~~~~,
\label{invariantm2}
\end{eqnarray}
Comparing to the $|M|^2$ from Eq.~(\ref{invariantm1}), the invariant matrix element squared for $q+\bar{q}\rightarrow g+\gamma^*$ is equivalent to that for $q+g\rightarrow q+\gamma^*$ with $p_2$ and $-p_5$, and $m_2$ and $m_5$ being exchanged and an additional overall minus sign.
The integration over phase space for the NLO processes is given by
\begin{eqnarray}
\int \frac{d^3p_5}{(2\pi)^3 2E_5}\int \frac{d^3p_3}{(2\pi)^3 2E_3}\int \frac{d^3p_4}{(2\pi)^3 2E_4}\nonumber\\
\times (2\pi)^4 \delta^{(4)}(p_1+p_2-p_3-p_4-p_5).
\end{eqnarray}

Introducing $q=p_3+p_4$, the phase space is factorized as following:
\begin{eqnarray}
\int \frac{d^3p_5}{(2\pi)^3 2E_5}\int d^4q \delta^{(4)}(p_1+p_2-q-p_5)\nonumber\\
\times\int \frac{d^3p_3}{(2\pi)^3 2E_3}\int \frac{d^3p_4}{(2\pi)^3 2E_4}(2\pi)^4 \delta^{(4)}(q-p_3-p_4),\nonumber\\
\label{phasesp}
\end{eqnarray}
where the second line is nothing but the phase space for a two-body decay, which affects only the dilepton part of Eq.~(\ref{dileptons}):

\begin{eqnarray}
\int \frac{d^3p_3}{(2\pi)^3 2E_3}\int \frac{d^3p_4}{(2\pi)^3 2E_4}(2\pi)^4 \delta^{(4)}(q-p_3-p_4)L^\beta L^{\beta^\prime*}\nonumber\\
=16\pi\alpha\int \frac{d^3p_3}{(2\pi)^3 2E_3}\int \frac{d^3p_4}{(2\pi)^3 2E_4}(2\pi)^4 \delta^{(4)}(q-p_3-p_4)\nonumber\\
\times \frac{p_3^\beta p_4^{\beta^\prime} +p_4^\beta p_3^{\beta^{\prime}} -g^{\beta{\beta^\prime}}q^2/2}{q^4}\nonumber\\
=16\pi\alpha\int \frac{d^3p_3}{(2\pi)^3 2E_3}\int \frac{d^3p_4}{(2\pi)^3 2E_4}(2\pi)^4 \delta^{(4)}(q-p_3-p_4)\nonumber\\
\times \frac{q^\beta q^{\beta^\prime} -2p_3^\beta p_3^{\beta^{\prime}} -g^{\beta{\beta^\prime}}q^2/2}{q^4}\nonumber\\
=16\pi\alpha\int \frac{d^3p_3}{(2\pi)^3 2E_3}\int \frac{d^3p_4}{(2\pi)^3 2E_4}(2\pi)^4 \delta^{(4)}(q-p_3-p_4)\nonumber\\
\times \frac{q^\beta q^{\beta^\prime} -2E_3^2 g^{\beta 0}g^{\beta^{\prime}0}-(2/3)|\vec{p_3}|^2g^{\beta i}g^{\beta^{\prime}i} -g^{\beta{\beta^\prime}}q^2/2}{q^4},\nonumber\\
\label{2body}
\end{eqnarray}
considering
\begin{eqnarray}
\int d^3p_3 p_3^i p_3^j =\frac{1}{3}\int d^3p_3 |\vec{p_3}|^2\delta^{ij}.
\end{eqnarray}
In the $q-$rest frame, Eq.~(\ref{2body}) reduces to
\begin{eqnarray}
\frac{16\pi\alpha}{q^4}\bigg\{q^\beta q^{\beta^\prime} -\frac{q^2}{2}(g^{\beta{\beta^\prime}}+g^{\beta 0}g^{\beta^{\prime}0}) -\frac{2|\vec{p_3}|^2}{3}g^{\beta i}g^{\beta^{\prime}i}\bigg\}\frac{|\vec{p_3}|}{4\pi q}\nonumber\\
\equiv \overline{L^\beta L^{\beta^\prime*}}\frac{|\vec{p_3}|}{4\pi q}=\overline{L^\beta L^{\beta^\prime*}}\frac{1}{8\pi}\sqrt{1-\frac{4 m_l^2}{q^2}},~~~~~
\label{phasesp1}
\end{eqnarray}
where $\overline{L^\beta L^{\beta^\prime*}}$ is the lepton pair tensor averaged over phase space.
The rest part of phase space, the first line in Eq.~(\ref{phasesp}), can be simplified as follows:
\begin{eqnarray}
\int \frac{d^3p_5}{(2\pi)^3 2E_5}\int d^4q \delta^{(4)}(p_1+p_2-q-p_5)\nonumber\\
=\frac{1}{(2\pi)^2}\int \frac{dp_5p_5^2 d\cos\theta}{2E_5}
=\frac{1}{16\pi^2\sqrt{s}}\int dq^2 d\cos\theta p_5\nonumber\\
=\frac{1}{16\pi^2\sqrt{s}}\int dq^2 d\cos\theta \sqrt{\frac{(s+m_5^2-q^2)^2}{4s}-m_5^2},\nonumber\\
\label{phasesp2}
\end{eqnarray}
with $p_1+p_2=p_5+q$ and $dq^2=d(p_1+p_2-p_5)^2=-2\sqrt{s}~dE_5$ in the center-of-mass frame.

Combining Eqs.~(\ref{phasesp1}) and (\ref{phasesp2}), the differential cross section is given by
\begin{eqnarray}
\frac{d\sigma}{dq^2d\cos\theta}=\frac{1}{8(4\pi)^3 p_i s}\sqrt{\frac{(s+m_5^2-q^2)^2}{4s}-m_5^2}\nonumber\\
\times\sqrt{1-\frac{4 m_l^2}{q^2}}\overline{|M|}^2,
\end{eqnarray}
with $L^\beta L^{\beta^\prime*}$ being substituted with $\overline{L^\beta L^{\beta^\prime*}}$ and $\overline{|M|}^2=|M|^2/96$ for $q+g\rightarrow q+\gamma^*$ and $\overline{|M|}^2=|M|^2/36$ for $q+\bar{q}\rightarrow g+\gamma^*$ from spin+color degeneracies; $p_i$ is the momentum of the initial particle in the center-of-mass frame:
\begin{eqnarray}
p_i=\sqrt{\frac{\{s-(m_1+m_2)^2\}\{s-(m_1-m_2)^2\}}{4s}}.
\end{eqnarray}
The numerical calculations are carried out in the $q$-rest frame:
\begin{eqnarray}
p_1^\mu&=&(E_1,~0,~0,|\vec{p_1}|),\nonumber\\
p_2^\mu&=&(E_2,~0,~|\vec{p_5}|\sin\psi,~|\vec{p_5}|\cos\psi -p_1),\nonumber\\
p_5^\mu&=&(E_5,~0,~|\vec{p_5}|\sin\psi,~|\vec{p_5}|\cos\psi),\nonumber\\
p_3^\mu&=&(E_3,~|\vec{p_3}|\sin\theta_1\sin\phi,~|\vec{p_3}|\sin\theta_1\cos\phi,~|\vec{p_3}|\cos\theta_1),\nonumber\\
p_4^\mu&=&(E_4,~-\vec{p_3}),
\end{eqnarray}
where
\begin{eqnarray}
|\vec{p_3}|&=&\sqrt{\frac{q^2}{4}-m_l^2},\nonumber\\
E_3&=&E_4=\sqrt{m_l^2+p_3^2},\nonumber\\
E_5&=&\frac{s-q^2-m_5^2}{2q^2}~~~ {\rm from}~s=(p_5+q)^2, \nonumber\\
|\vec{p_5}|&=&\sqrt{E_5-m_5^2},\nonumber\\
E_2&=&\frac{m_2^2+q^2-t}{4E_3}~~~ {\rm from}~t=(p_2-q)^2, \nonumber\\
E_1&=&\sqrt{s+|\vec{p_5}|^2}-E_2~~~ {\rm from}~s=(p_1+p_2)^2, \nonumber\\
\cos\psi&=&\frac{m_2^2+|\vec{p_1}|^2 +|\vec{p_5}|^2 -E_2^2}{2 |\vec{p_1}||\vec{p_5}|}\nonumber
\end{eqnarray}
from $E_2^2-|\vec{p_5}|^2\sin^2\psi-(|\vec{p_5}|\cos\psi-p_1)^2=m_2^2$.
Independent variables are then $s$, $t$, $\theta_1$, and $\phi$. Integrating over $\theta_1$ and $\phi$ in the $q$-rest frame we get

\begin{eqnarray}
\overline{L^\beta L^{\beta^\prime*}}=16\pi\alpha\frac{(q^2/2) -(2/3)|\vec{p_3}|^2}{q^4}g^{\beta i}g^{\beta^{\prime}i}\nonumber\\
=16\pi\alpha\frac{q^2 -2m_l^2}{3q^4}g^{\beta i}g^{\beta^{\prime}i}
\end{eqnarray}
from Eq. (\ref{phasesp1}).


\hfil\break
\appendix
\centerline{\bf \large Appendix B}
\bigskip

In this appendix we study the effect of acceptance cuts on the invariant mass spectrum of dielectrons and  the dependence of the dielectron spectrum on the total cross section for charm production (within the experimental uncertainties) by considering minimal-bias Au+Au collisions at $\sqrt{s_{\rm NN}}$ = 200 GeV (as an example).

\begin{figure} [h!]
\centerline{
\includegraphics[width=8.6 cm]{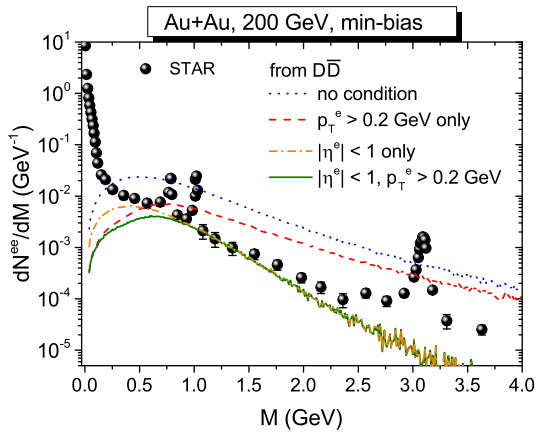}}
\caption{The invariant mass spectra of dielectrons from $D\bar{D}$ pairs in minimum-bias Au+Au collisions at $\sqrt{s_{\rm NN}}$ = 200 GeV without any acceptance cuts (dotted blue line), only with the $p_T$ cut
(dashed red line), only with the pseudo-rapidity cut $|\eta^e| <$ 1 (dot-dashed orange line), and with both cuts included (solid green line). The full dots are the experimental data from the STAR Collaboration for the total dilepton spectra.}
\label{app2a}
\end{figure}

Fig.~\ref{app2a} shows the invariant mass spectra of dielectrons from $D\bar{D}$ pairs with several acceptance cuts incorporated.
The dotted blue line is the dielectron spectrum without any acceptance cuts and is naturally much higher than the data.
The dashed red line is the spectrum with the $p_T$ cut for electrons as well as  positrons ($|p_T^e| >$  0.2 GeV).
This cut reduces the dielectron mass spectrum slightly more at low invariant mass than at large invariant masses and thus enhances the apparent slope for intermediate masses. This results from the fact that
the electron and the positron with large invariant mass have large momenta such that the $p_T^e$ cut is less effective.
On the other hand, the pseudo-rapidity cut ($|\eta^e| <$ 1) reduces considerably the dielectron spectrum at large invariant mass for the same reason.
If the momenta of electron and positron - composing a dielectron - are large due to the large invariant mass, they tend to lie outside the pseudo-rapidity cut.
The solid green line, finally,  is the dielectron spectrum after both cuts, which is essentially the same as the green line in Fig.~\ref{fig8} (e).

\begin{figure} [h!]
\centerline{
\includegraphics[width=8.6 cm]{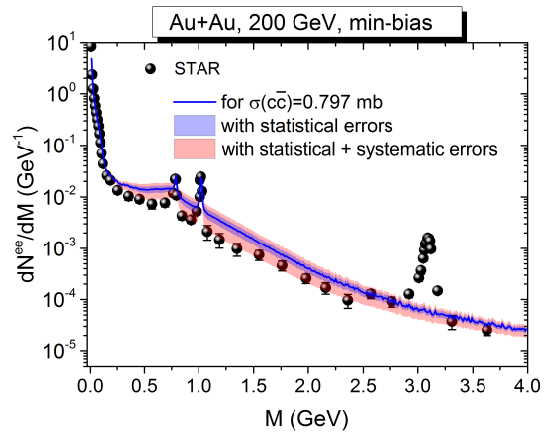}}
\caption{The invariant mass spectra of dielectrons for minimum-bias Au+Au collisions at $\sqrt{s_{\rm NN}}$ = 200 GeV with the total cross section for charm production from the STAR collaboration with statistical and systematic errors displayed in terms of the shaded areas.}
\label{app2b}
\end{figure}

According to the STAR measurements, the total cross section for charm production in p+p collisions at $\sqrt{s_{\rm NN}}$ = 200 GeV is $797\pm 210^{+208}_{-295}~{\rm \mu b}$~\cite{Adamczyk:2012af} and thus has a statistical and systematic error about a factor of two.
Fig.~\ref{app2b} shows the invariant mass spectrum of dielectrons with the charm cross section from the STAR collaboration considering its statistical and systematic errors while including the cuts in $p_T^e$ and $\eta^e$.
Since the dielectrons from $D\bar{D}$ semi-leptonic decays are the most dominant contribution in the intermediate mass range, the total dielectron spectrum is primarily sensitive to the charm cross section employed.
The figure shows that the PHSD results  with the mean value of charm cross section (from STAR) overestimate
the dielectron data and the inclusion of both statistical and systematic errors is necessary to achieve an agreement with the experimental data.
For our present study we use the charm cross sections fitted to the experimental data within a wide range of collision energies as shown in Fig.~\ref{fig1} (a), where the cross section at
$\sqrt{s_{\rm NN}}$ = 200 GeV is about 400 ${\rm \mu b}$, which is still within the statistical and systematic error bars of the STAR collaboration and which is close to the recent results from the PHENIX collaboration \cite{Adare:2017caq}.

\end{document}